\documentclass[%
reprint,
superscriptaddress,
 amsmath,amssymb,
 aps,
 prl,
showkeys]{revtex4-2}

\usepackage{graphicx}
\usepackage[caption=false]{subfig}
\usepackage{dcolumn}
\usepackage{bm}
\usepackage{mathtools}
\usepackage{xcolor,soul}
\usepackage{chngcntr}
\usepackage{multirow}
\usepackage{tablefootnote} 
\usepackage{booktabs}
\usepackage{multirow}
\usepackage[unicode=true]{hyperref}
\usepackage{color}
\usepackage{makecell}
\usepackage{url}
\usepackage{gensymb}
\usepackage{booktabs}

\newcommand{\ls}[4]{\ensuremath{^{#1} {#2}_{#3}^{#4}}}
\newcommand{\NIST}{
National Institute of Standards and Technology, 325 Broadway, Boulder, Colorado 80305, USA}
\newcommand{\CU}{
Department of Physics, University of Colorado, Boulder, Colorado 80309, USA}

\begin{document}

\preprint{}

\title{Cryogenic Optical Lattice Clock with $1.7\times 10^{-20}$ Blackbody Radiation Stark Uncertainty}%

\affiliation{\NIST}
\affiliation{\CU}
\affiliation{Electrical, Computer \& Energy Engineering, University of Colorado, Boulder, Colorado 80309, USA}
\affiliation{National Metrology Institute of Japan (NMIJ), National Institute of Advanced Industrial Science and Technology (AIST), 1-1-1 Umezono, Tsukuba, Ibaraki 305-8563, Japan}

\author{Youssef S. Hassan}
\affiliation{\NIST}
\affiliation{\CU}

\author{Kyle Beloy}
\affiliation{\NIST}

\author{Jacob L. Siegel}
\affiliation{\NIST}
\affiliation{\CU}

\author{Takumi Kobayashi}
\affiliation{\NIST}
\affiliation{National Metrology Institute of Japan (NMIJ), National Institute of Advanced Industrial Science and Technology (AIST), 1-1-1 Umezono, Tsukuba, Ibaraki 305-8563, Japan}

\author{Eric Swiler}
\affiliation{\NIST}
\affiliation{\CU}

\author{Tanner Grogan}
\affiliation{\NIST}
\affiliation{\CU}

\author{Roger C. Brown}
\affiliation{\NIST}
\affiliation{\CU}

\author{Tristan Rojo}
\affiliation{\NIST}
\affiliation{\CU}

\author{Tobias Bothwell}
\affiliation{\NIST}

\author{Benjamin D. Hunt}
\affiliation{\NIST}
\affiliation{\CU}

\author{Adam Halaoui}
\affiliation{\NIST}

\author{Andrew D. Ludlow}
\email{andrew.ludlow@nist.gov}
\affiliation{\NIST}
\affiliation{\CU}
\affiliation{Electrical, Computer \& Energy Engineering, University of Colorado, Boulder, Colorado 80309, USA}


\date{\today}
\begin{abstract}
Controlling the Stark perturbation from ambient thermal radiation is key to advancing the performance of many atomic frequency standards, including state-of-the-art optical lattice clocks (OLCs). We demonstrate a cryogenic OLC that utilizes a dynamically actuated radiation shield to control the perturbation at $1.7\times10^{-20}$ fractional frequency, a factor of $\sim$40 beyond the best OLC to date. Our shield furnishes the atoms with a near-ideal cryogenic blackbody radiation (BBR) environment by rejecting external thermal radiation at the part-per-million level during clock spectroscopy, overcoming a key limitation with previous cryogenic BBR control solutions in OLCs. While the lowest BBR shift uncertainty is realized with cryogenic operation, we further exploit the radiation control that the shield offers over a wide range of temperatures to directly measure and verify the leading BBR Stark dynamic correction coefficient for ytterbium. This independent measurement reduces the literature-combined uncertainty of this coefficient by 30\%, thus benefiting state-of-the-art Yb OLCs operated at room temperature. We verify the static BBR coefficient for Yb at the low $10^{-18}$ level.
\end{abstract}


\maketitle

Thermal radiation is ubiquitous in nature, inescapable even in outer space. An ideal blackbody radiation (BBR) environment---characteristic of isothermal surroundings---is homogeneous, isotropic, spectrally Planckian, and characterized exclusively by its temperature. In atomic clocks, this thermal radiation perturbs the atomic energy levels, inducing a Stark shift to the clock frequency conventionally known as the BBR shift. For over four decades, the BBR shift has been a persistent challenge in the pursuit of better atomic clock performance~\cite{ItaLewWin82,LudBoyYe15}.

For room-temperature optical lattice clocks (OLCs) based on Yb and Sr, the BBR shift represents the largest uncanceled systematic frequency shift, amounting to $-2.4\times10^{-15}$ and $-4.8\times10^{-15}$ \cite{Note1}, respectively, while typically also constituting the largest source of uncertainty in these clocks~\cite{McGZhaFas18,AepKimWar24}. The uncertainty in the BBR shift stems from (i) uncertainty in the BBR environment, specific to the apparatus and assessment techniques, and (ii) uncertainty in the atomic response to BBR, specific to the atomic species. For evaluation purposes, the BBR shift is partitioned into a dominant ``static'' contribution and a smaller ``dynamic'' correction \cite{BelHinPhi14}. BBR shift uncertainties as low as $\approx1\times10^{-18}$ have been reported for room-temperature Yb and Sr OLCs with the leading uncertainty attributed to the dynamic correction of the atomic response ~\cite{SheLemHin12,MidFalLis12,safronova2012ytterbium,BelSheLem12,safronova2013blackbody,BelHinPhi14,LisDorNos21,heo2022evaluation,AepKimWar24}. This motivates independent evaluations of the dynamic correction using varied approaches. At the same time, an appealing avenue towards better clock accuracy is to eliminate the need for precise knowledge of the dynamic correction altogether.

At cryogenic temperatures, the BBR shift is highly suppressed. Of more practical significance, however, is that (i) the temperature sensitivity is suppressed, providing greater tolerance to absolute temperature uncertainty and (ii) the uncertainty due to the atomic response is rendered negligible ($<\!10^{-21}$) for both Yb and Sr. Despite these clear benefits, cryogenic operation is hindered by the practical needs of clock operation, which include preparing and optically addressing lattice-trapped atoms while maintaining a well-characterized BBR environment. 

Cryogenic OLCs have previously been demonstrated in Refs.~\cite{UshTakDas15,takamoto_frequency_2015,Sch22}. With two distinct designs, the interrogated atoms are predominantly surrounded by cryogenic surfaces, suppressing the BBR shift. One design relies on shuttling lattice-trapped atoms to a small cryogenic enclosure for interrogation with reduced exposure to external thermal radiation. The other design utilizes a double-layered static shield in which the lattice-trapped atoms are loaded and interrogated. In both designs, however, the atoms retain direct line of sight to external thermal radiation, which contributes significantly to the shift. Residual factors such as reflections within the cryogenic region or transmission through windows must be precisely taken into account. Ultimately, uncertainties associated with these factors lead to BBR shift uncertainties in those cryogenic OLCs that are comparable to what has been achieved with room-temperature OLCs ($\approx 1\times10^{-18}$).

\begin{figure*}
    \centering
    \includegraphics[width=1\linewidth]{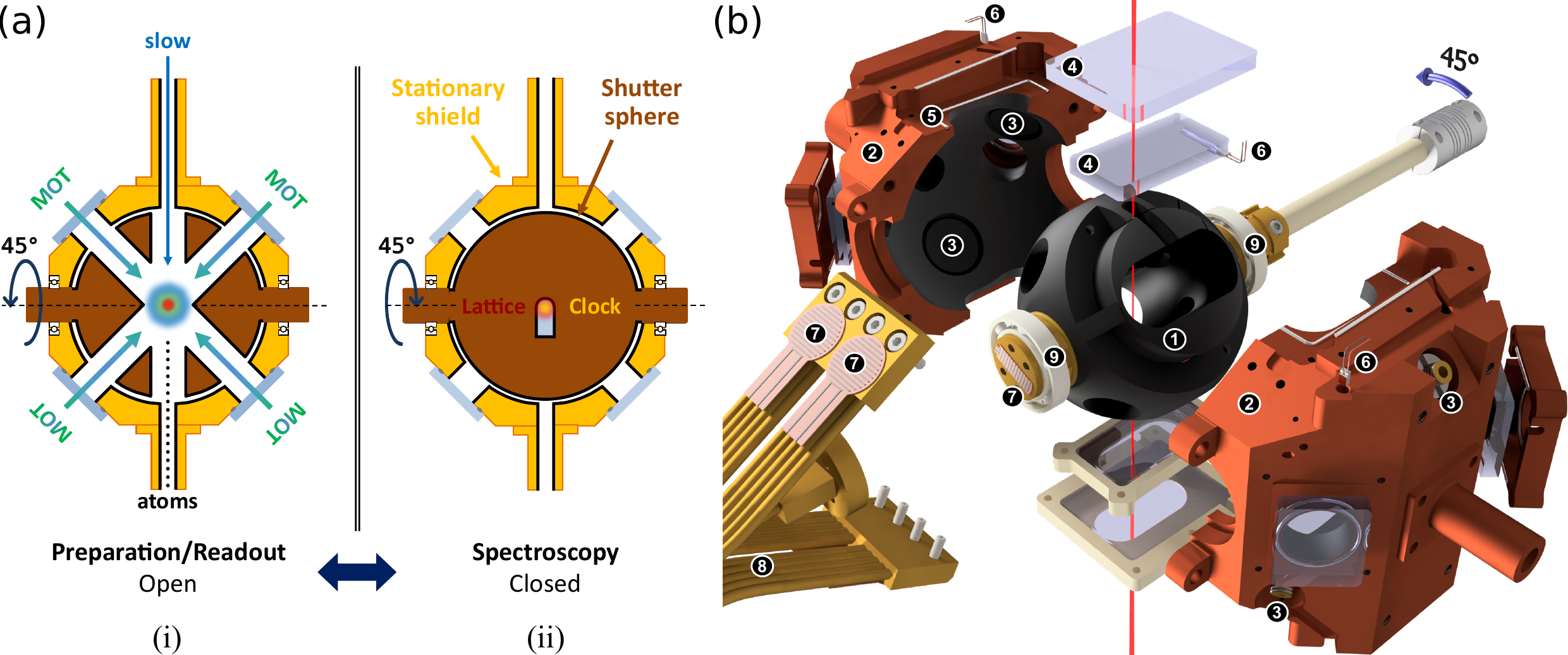}
    \caption{Illustrations of the shield structure and function. (a) Horizontal cross section at mid-plane of the shield. (i) Open configuration when the apertures of the shutter sphere (brown) and the stationary shield (gold) are aligned, providing physical and optical access for the atomic beam and the magneto-optical trap (MOT) beams to cool and trap the atoms. After loading the atoms into the vertical optical lattice, the shutter sphere is rotated around the dashed axis, shuttering direct exposure to external thermal radiation. (ii) Closed configuration when clock spectroscopy is performed in a near-ideal BBR environment determined by the controlled shield temperature. A slit on the shutter sphere maintains optical access for the vertical lattice and the co-propagating clock spectroscopy beam. (b) Exploded 3D rendering of the shield. The shutter sphere (1), 6~cm in diameter, is shown in the closed configuration for spectroscopy, with its apertures facing the internal surface of the stationary radiation shield (2) and aligned with electrodes for dc Stark evaluation (3). The lattice beam (vertical red line) is permitted through the slit on the sphere, while external thermal radiation is blocked by the double stack of absorptive windows (4) thermally coupled to the shield through pressed indium (5). RTDs are distributed (6) to measure the temperature of the sphere (3$\times$), the two stationary shield halves (4$\times$), and a vertical window with direct line of sight to the atoms during spectroscopy (1$\times$) (only 3 RTDs are shown). Three heater sets (7) control and stabilize the shield temperature. Flexible copper braids (8) to the cryocooler tip accommodate the sphere rotation. All internal copper surfaces are coated with absorptive black coating. Ceramic ball bearings (9) support the sphere inside the shield. An out-of-vacuum stepper motor rotates the sphere between closed and open configurations in $\approx$100 ms with minimal vibration owing to its balanced design.}
    \label{fig:Shield}
\end{figure*}

In this Letter, we demonstrate a cryogenic \ls{171}{\mathrm{Yb}}{}{} OLC, achieving a BBR shift uncertainty of $1.7\times10^{-20}$. Our core strategy is to furnish the interrogated atoms with a near-ideal BBR environment at a cryogenic temperature. Below we describe a BBR enclosure, or ``shield,'' that accomplishes this task by dynamically shuttering all but the necessary optical access for clock spectroscopy. With this design, external thermal radiation is thoroughly extinguished, with its small residual effects bounded through radiation modeling. Conservative uncertainty estimates of the resulting residual shift are then assigned with minimal reliance on the details of the model. Instead, the leading uncertainty is attributed to thermal gradients across the shield, which are evaluated with contact thermometers embedded in the shield components. While the BBR shift uncertainty is minimized at cryogenic temperatures, the shield is also capable of providing near-ideal BBR environments over a wide range of temperatures, from $\approx77$~K to $\approx320$~K. This capability facilitates, for the first time in Yb, an independent determination of the leading dynamic BBR correction by direct observation, achieving a level of precision comparable to other methods. Importantly, while our shield is implemented in an Yb OLC, the design is general to all OLC species.

An illustration of the basic function of the shield is shown in Fig.~\ref{fig:Shield}a. The shield consists of a radiation enclosure with a sphere-shaped core, referred to as the ``the shutter sphere," that is actuated between an ``open'' and a ``closed" configuration. The open configuration provides access for atomic sample preparation and readout. The closed configuration provides a controlled BBR environment for clock spectroscopy. In the closed configuration, the atoms have no direct exposure to the external environment and are exclusively surrounded by highly emissive surfaces whose temperatures are monitored and controlled in real time. A small vertical optical access is maintained for the lattice and the co-propagating clock spectroscopy beam. This optical access---subtending $<$1\% of the solid angle around the atoms---is fitted with double-stacked N-BK7 substrates that block external thermal radiation and are thermally coupled to the shield.

The detailed components of the shield are shown in the 3D exploded rendering in Fig.~\ref{fig:Shield}b. The main body of the shield, machined from copper for high thermal homogeneity, consists of two halves enclosing the shutter sphere. The shutter sphere is mechanically rotated to selectively block all but the optical access required for clock spectroscopy, realizing the open and closed configurations of the shield. The entire assembly is mounted within an ultra-high vacuum chamber and cooled through a vibration-isolating helium flow cryocooler \cite{commercial}. To accommodate the mechanical rotation of the shutter sphere, the cold tip is thermally connected to the shield through a group of flexible copper braids.

Several peripheral components and features are critical for the shield's function in maintaining and characterizing its BBR environment. Eight calibrated negative-temperature-coefficient resistance temperature detectors (RTDs) \cite{commercial} are embedded in strategic locations in the shield as detailed in Fig.~\ref{fig:Shield}. These RTDs provide real-time measurement of the absolute temperature and thermal gradients of the shield surfaces that have a direct line of sight to the atoms during spectroscopy. Additionally, a heater film set is attached to each of the three copper body components (shutter sphere and shield halves) to compensate for imbalances between the different thermal loads and cooling powers to which they are exposed. Each set regulates the mean temperature of the attached component to a programmable set-point via an independent feedback loop, providing a uniform temperature enclosure for the atoms. 

The internal copper surfaces of the shield are coated with an electrically dissipative black coating, contributing to a Faraday enclosure that protects the atoms during spectroscopy from possible stray-charge-induced dc Stark shifts. The apertures on the stationary shield intended for optical access are fitted with N-BK7 optical substrates (pale blue in Fig.~\ref{fig:Shield}a), which are largely opaque to the external thermal radiation. The apertures for the atomic thermal beam are equipped with high-aspect-ratio tubes to reduce external radiation influx. Both the substrates and the tubes are thermally coupled to the shield through pressed indium, and contribute to the overall external radiation suppression.

A core feature of the shield is its actuated nested design, which effectively attenuates external radiation entering through the apertures. This attenuation occurs through multiple reflections off the highly absorptive black coatings and through the absorption of radiation by the window substrates. As a result, only residual amounts of the radiation ultimately reach the atoms. At 77~K, we estimate an upper bound of $4\times10^{-21}$ on the shift caused by the residual external radiation from all apertures. This assessment is based on a reverse ray-tracing modeling with conservative estimates of both the emissivity of the black coatings and the transmission properties of the window over the whole electromagnetic spectrum \cite{SM}. We assign this upper bound as the uncertainty for the residual radiation shift. 

Another critical feature is the high thermal homogeneity maintained by the internal shield surfaces during spectroscopy, achievable across a wide temperature range from 77 K to 318 K. Despite being constructed from three separate components of bulk copper, setting the mean temperature of each component to a common temperature leads to an overall BBR homogeneity limited by the copper thermal conductivity and the RTD calibration uncertainty. The shield achieves stability and in-loop temperature control at the sub-millikelvin level for averaging times $>8$~s for each component, facilitated by real-time thermal feedback \cite{SM}. We utilize the RTD measurements, corroborated with thermal simulations, to capture the thermal extremes of all internal surfaces of the shield with direct line of sight to the atoms \cite{SM}. Weighted by the solid angles subtended by those surfaces, we use those temperature extremes to estimate a uniform probability distribution of the induced shift at the atoms. We then assign its standard deviation ($1.6\times 10^{-20}$) as the uncertainty on the shift, corresponding to a BBR temperature uncertainty of 29~mK. We perform similar analysis at 298.5~K and 318.2~K. The remaining sources of uncertainty of the BBR shift are detailed in Table~\ref{Tab:errorbudget} and in \cite{SM}.

\begin{table}
\caption{BBR shift uncertainty ($\times10^{-21}$ clock frequency) for Yb2 for three typical mean set temperatures. DLS: direct-line-of-sight.}
\label{Tab:errorbudget}
\begin{ruledtabular}
\begin{tabular}{lccc}
\multicolumn{1}{c}{{\it BBR environment}} &\multicolumn{1}{c}{{77.0 K}}&\multicolumn{1}{c}{{298.5 K}}&\multicolumn{1}{c}{{318.2 K}}\\
RTD temperature measurements						\\
\quad Manufacturer calibration			   	& $3.4$ & $580$& $810$  \\
\quad Electronic measurement error		    & $2.2$ & $130$ & $160$ \\
\quad Self-heating					        & $<0.1$ & $1$& $1$ \\
\quad Parasitic conduction/radiation		& $1$  & $<1$ & $<1$ \\
Temperature inhomogeneity \\
\quad DLS surfaces                          & $16$  & $130$ & $320$ \\
\quad Residual external radiation            & $3$   & $<1$  & $5$ \\
\quad Non-DLS surfaces \& others            & $4$   & $2$   & $30$ \\
\hline
\multicolumn{1}{c}{{\it atomic response}} &\multicolumn{1}{c}{}&\multicolumn{1}{c}{}&\multicolumn{1}{c}{}\\
Static polarizability		             	& $0.2$  & $45$ & $60$ \\
Dynamic corrections             			& $0.3$  & $940$ & $1380$ \\
BBR Zeeman factor					& $<0.01$ & $<1$ & $<1$ \\
\hline
Total BBR environment			        	& $17$ & $600$ & $890$\\
Total atomic response			        	& $0.4$ & $940$ & $1380$\\
Total 							           	& $17$ & $1120$ & $1640$
\end{tabular}
\end{ruledtabular}
\end{table}

Electrical insulation is placed between the shield halves to suppress eddy currents during the MOT loading phase, produced by large changes in the vertical magnetic flux. A thin vertical cut in the shutter sphere serves the same function. Three mutually perpendicular pairs of electrodes, thermally coupled to but electrically insulated from the shield, are positioned to have a direct line of sight to the atoms during spectroscopy. The electrodes, grounded during clock operation, are utilized in dedicated measurements to verify the absence of a dc Stark shift, for example, due to the accumulation of stray static charges. 

\begin{figure}[b]
\centering
\includegraphics[width=1.0\columnwidth]{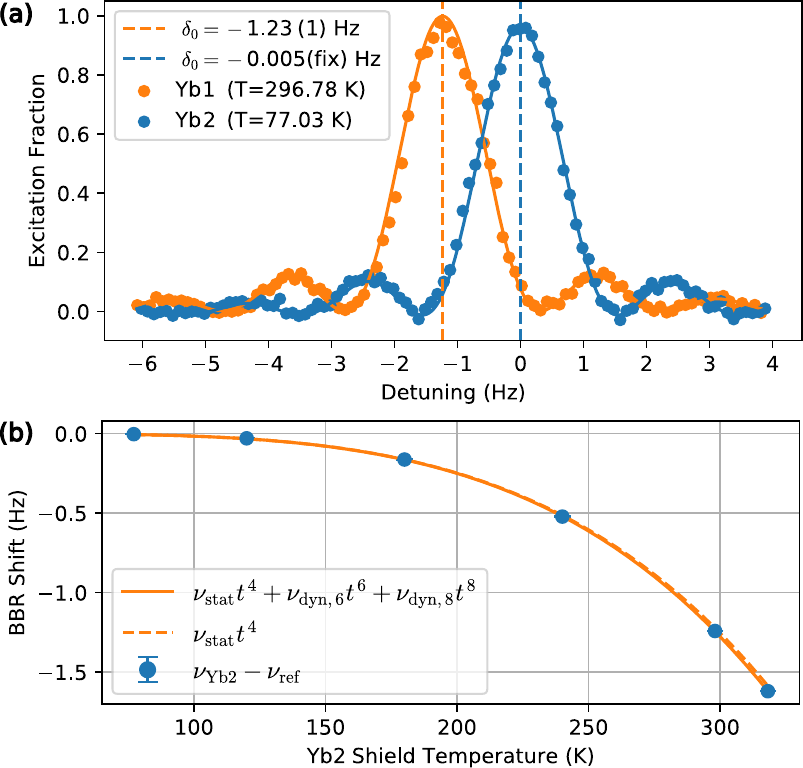}
\caption{(a) Average of two simultaneous Rabi line scans on two Yb clocks with the Yb2 shield at cryogenic temperature and the Yb1 shield at room temperature. The solid lines are Rabi lineshape fits with center frequency $\delta_0$. Both scans have been corrected for first order magnetic shifts at the 15~mHz level, with all other corrections well below that. Zero detuning is defined at zero BBR shift relative to Yb2 center frequency. (b) BBR shift of Yb2 as a function of the shield temperature as compared to the reference clock. Both clock frequencies are corrected for all known systematic shifts, except for the BBR shift on Yb2. The solid orange line is the best fit of Eq.~(\ref{Eq:nuBBR}) with fixed dynamic coefficients. The dashed line represents the static contribution to the BBR shift. The 68\% confidence error bars are small compared to the total BBR shift scale.}
\label{fig:line scan}
\end{figure}

The cryogenic shield is installed in one of our Yb lattice clock systems, referred to as Yb2. In contrast, our other Yb clocks, Yb1 and YbT, employ room-temperature BBR control solutions. These serve as stable frequency references in the measurements described below. Specifically, Yb1, a laboratory-based clock, utilizes an in-vacuum shield at room temperature that is passively coupled to the vacuum chamber \cite{BelSheLem12}. Meanwhile, YbT, a transportable clock, features a small, temperature-stabilized vacuum chamber that functions as its BBR shield \cite{PortInPrep}. The three systems share the clock laser, while Yb1 and Yb2 share cooling and lattice lasers. Yb2 alone utilizes an optical enhancement cavity for the lattice. 

We first observe the total BBR shift of the Yb clock transition due to room-temperature radiation by performing simultaneous narrow-line spectroscopy with a 560~ms Rabi pulse on Yb2 and Yb1. The Yb2 shield temperature is set at 77.03 K, while the Yb1 shield is at 296.78~K. Line scans and fits shown in Fig.~\ref{fig:line scan}a reveal a clock shift that is consistent at the $1\sigma$ level of the expected BBR shift after accounting for the first-order Zeeman shift in such scans. 

By varying the temperature of the Yb2 shield, we apply the BBR shift model 
\begin{equation}
    \nu_\mathrm{BBR}(t)=\nu_\mathrm{stat}t^4+\nu_\mathrm{dyn,6}t^6+\nu_\mathrm{dyn,8}t^8+ \mathcal{O}(t^{10}),
\label{Eq:nuBBR}
\end{equation}
to the clock frequencies at five selected Yb2 shield temperatures. Here $t \equiv T/T_0$ is a normalized temperature parameter for a BBR environment characterized by an absolute temperature $T$, with conventional reference temperature $T_0 = 300$~K. The first term on the right-hand side of Eq.~(\ref{Eq:nuBBR}) is the static approximation to the BBR shift, while the remaining terms are dynamic corrections that account for the spectral dependence of the atomic polarizabilities and the BBR. 

In Fig.~\ref{fig:line scan}b, the Yb2 frequency difference from a reference Yb clock ($\nu_\mathrm{Yb2}-\nu_\mathrm{ref}$) is plotted as a function of the shield temperature. In this case, the reference Yb clock is either Yb1 or YbT, both corrected for all systematics, while Yb2 is corrected for all systematics except for the BBR shift. Equation~(\ref{Eq:nuBBR}) is then fit to measurements of the frequency difference at six different temperatures, in this case fixing $\nu_\mathrm{dyn,6}$ and $\nu_\mathrm{dyn,8}$ to the established values in the literature \cite{safronova2012ytterbium,BelSheLem12}. The best fit is achieved with the static coefficient $\nu_\mathrm{stat}=-1.2545(10)\ \mathrm{Hz}$. This is in excellent agreement with the more precise measurement by application of a static electric field in Ref.~\cite{sherman2012high} and is self-consistent with the BBR corrections applied to Yb1 and YbT. We note that the fitted value of $\nu_\mathrm{stat}$ maintains good agreement with Ref.~\cite{sherman2012high}, regardless of the choice of $\nu_\mathrm{dyn,6}$, whether it be the value in Ref.~\cite{BelSheLem12} or the value reported later in this Letter. The measurement of $\nu_\mathrm{stat}$ here provides a confirmation of the dc differential polarizability of the \ls{171}{\mathrm{Yb}}{}{} clock transition, establishing an independent verification of the room-temperature static BBR correction at the $2\times 10^{-18}$ level.

\begin{figure}[b]
\centering
\includegraphics[width=1.0\columnwidth]{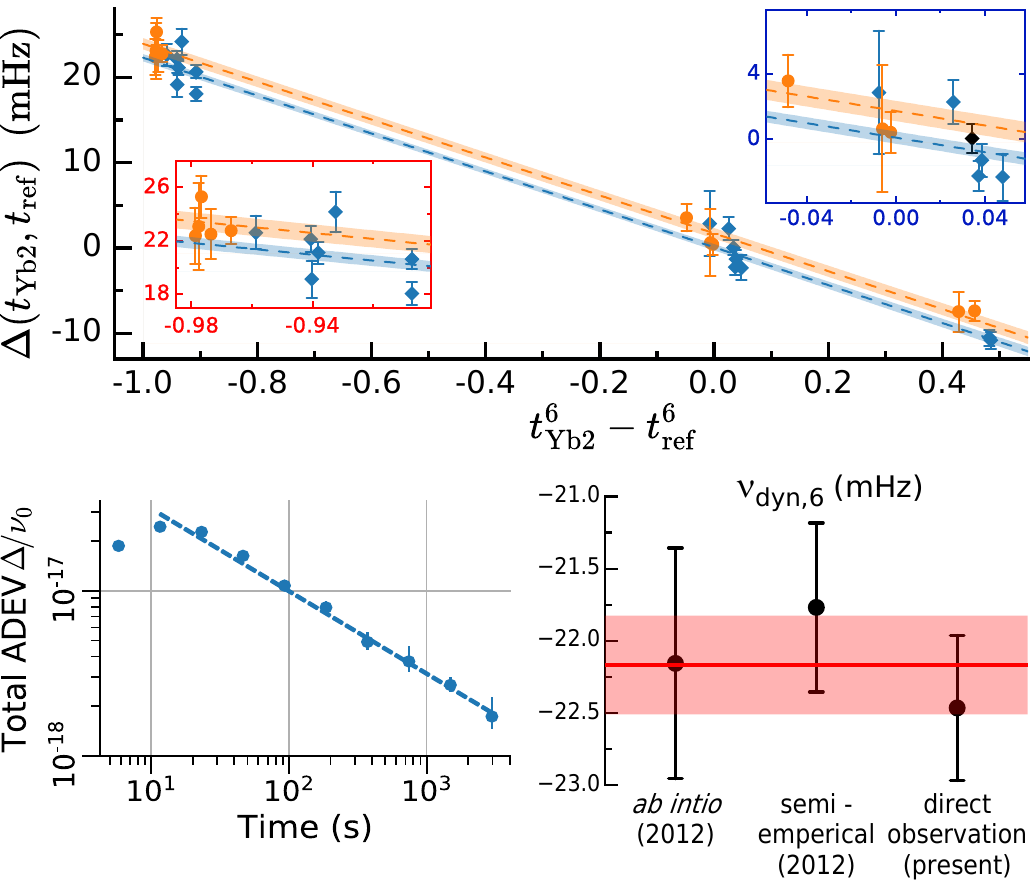}
\caption{(Top) Observed frequency difference between two pairs of Yb clocks: $\Delta(t_\mathrm{Yb2},t_\mathrm{Yb1})$ (blue) and $\Delta(t_\mathrm{Yb2},t_\mathrm{YbT})$ (orange). Vertical error bars represent combined statistical and differential systematic uncertainties, calculated as explained in the main text. The dashed lines and shaded bands represent the best fit line with slope $\nu_\mathrm{dyn,6}=-22.47(50)$~mHz. Insets are magnified scales of runs at 77~K (red frame) and 298~K (blue frame) Yb2 shield temperatures. (Bottom left) Total Allan deviation of a single synchronized run (black diamond in the top sub-figure inset) with an estimated instability of $1.10(3)\times10^{-18}$ fractional frequency at the full measurement time (8000~s). (Bottom right) Comparison between \textit{ab initio} \cite{safronova2012ytterbium}, semi-empirical \cite{BelSheLem12}, and direct experimental measurement (present work) of $\nu_\mathrm{dyn,6}$. The red horizontal line represents the weighted mean of the three values, with shaded error band representing the 68\% confidence interval.}
\label{fig:vdyn fit}
\end{figure}

Armed with the uniform BBR environment provided by the shield from cryogenic to above-room temperatures, we analyze a series of comparisons between the three clocks \cite{CompInPrep} to determine $\nu_\mathrm{dyn,6}$. We fit the observed frequency difference, corrected for all systematics except for the first-order dynamic contribution, to a linear model whose slope is $\nu_\mathrm{dyn,6}$ expressed as
\begin{equation}
    \Delta(t_\mathrm{Yb2},t_\mathrm{ref})=\nu_\mathrm{dyn,6}(t_\mathrm{Yb2}^6-t_\mathrm{ref}^6)+\nu_c,
\label{Eq:DeltanuBBR}
\end{equation}
where $t_\mathrm{Yb2}\ (t_\mathrm{ref})$ is the normalized temperatures of Yb2 (Yb1/YbT) and $\nu_c$ accommodates any frequency offset between clock pairs. Most measurements were conducted at Yb2 shield temperatures of 77 K, 298 K, and 318 K, while the reference clocks were close to room temperature. 

We performed a global fit to two expanded datasets of 15 measurements of $\Delta(t_\mathrm{Yb2},t_\mathrm{Yb1})$ and 10 measurements of $\Delta(t_\mathrm{Yb2},t_\mathrm{YbT})$ with a common slope and independent offsets, as shown in Fig.~\ref{fig:vdyn fit}. The best fit is achieved with $\nu_\mathrm{dyn,6}=-22.47(50)$~mHz. The reported uncertainty on the fit is inflated by the square root of the reduced chi squared $\sqrt{\chi^2_\mathrm{red}}=\sqrt{1.4}$. The statistical uncertainty for each measurement was determined by the total Allan deviation extrapolating a white noise model to the full measurement time \cite{noisefloorpts}. For about half of the measurements, we interrogated the clocks synchronously to reject common-mode clock laser noise and achieve better differential stability. The total Allan deviation plot of a selected synchronized measurement is also shown, with a white noise fit of $9.9(3)\times10^{-17}/\sqrt{\tau}$ fractional frequency instability where $\tau$ is the averaging time in seconds. For each measurement, we excluded from the differential systematic uncertainty any effects that do not influence the slope. Temperature drift and measurement uncertainty on the abscissa are negligible. However, we investigated temperature-dependent factors, such as the variation of the background gas pressure, which could bias the fitted $\nu_\mathrm{dyn,6}$ through a correlated background gas collisional shift. The reported uncertainty is inflated to account for the possible bias, as detailed in \cite{SM}. We note that any plausible background gas collisional shift bias is bounded to less than 0.6$\sigma$ from the reported measurement.

The fitted $\nu_\mathrm{dyn,6}=-22.47(50)$~mHz in this work is in good agreement with the \textit{ab initio} $\nu_\mathrm{dyn,6}=-22.16(80)$~mHz \cite{safronova2012ytterbium} and semi-empirical $\nu_\mathrm{dyn,6}=-21.77(59)$~mHz \cite{BelSheLem12} previously reported in the literature for Yb using entirely different methods \cite{vdyn6-8}. We thus report the weighted mean of the three determinations at $\nu_\mathrm{dyn,6}=-22.17(34)$~mHz, with 30\% reduced uncertainty compared to the combined uncertainty of the previous two values.

In summary, we describe the design and operation of a cryogenic radiation shield that realizes an unprecedented BBR Stark shift uncertainty of $1.7\times10^{-20}$ in an Yb OLC. The shield realizes such control by thoroughly rejecting external thermal radiation and enclosing the clock atoms from all $4\pi$ steradians by emissive and thermally uniform surfaces, creating a near-ideal cryogenic BBR environment during clock spectroscopy. Additionally, by modulating the shield temperature, we directly measure the BBR static and leading dynamic coefficients at the $2\times10^{-18}$ and $1\times10^{-18}$ levels, respectively. This reduces the latter's literature-combined uncertainty by 30\%, an improvement that is particularly valuable for enhancing the accuracy of Yb OLCs operated at room temperature.

\begin{acknowledgments}
We thank Jeffrey A. Sherman and Vladislav Gerginov for their careful reading of the manuscript and their critical feedback. We also thank Kyungtae Kim, Dahyeon Lee, and Jun Ye for providing cryogenic silicon cavity stable reference \cite{oelker2019demonstration} to our clock laser during a subset of the measurements in this work, as well as Nicholas V. Nardelli and Tara Fortier for operating the optical frequency comb that enabled this stability transfer. We wish to express our gratitude to Elizabeth A. Donley for her service as the NIST Time and Frequency Division Chief over the past several years. We gratefully acknowledge funding support from NIST, the Office of Naval Research, National Science Foundation QLCI Grant No. 2016244, and by an appointment to the Intelligence Community Postdoctoral Research Fellowship Program at NIST administered by ORISE through an inter-agency agreement between the U.S. DoE and the Office of the Director of National Intelligence. The authors declare no competing interests. Data supporting the findings of this study are available from the corresponding author on a reasonable request.

YSH led the design, assembly, and integration of the shield into Yb2, as well as its operation. KB and ADL contributed to the design. YSH, KB, and JLS jointly conducted the analysis. TK and KB contributed to the shield integration into the clock. TG was responsible for the maintenance of the optical local oscillators. The operation of Yb1 was handled by JLS and BDH, while EWS, RCB, TR, TB, and AIH operated YbT. The project was conceptualized by KB and ADL, who also supervised the entire project. All authors contributed to the final manuscript and discussed the results and the analysis.
\end{acknowledgments}

\section{End Matter}
As mentioned in the main text, the shield functions as a Faraday enclosure around the atoms. To verify the absence of any dc Stark shift from stray electric fields, we utilize three pairs of electrodes exposed to the atoms in the closed configuration. These electrodes are arranged to generate controlled electric fields in three nominally mutually orthogonal directions to assess and constrain the dc Stark shift at the atoms' location. Following \cite{BelZhaMcG18}, we constrain the dc Stark shift in the grounded configuration to $<10^{-19}$ \cite{SM}. Tighter constraints should be readily attainable, if deemed necessary for future clock operation.

\clearpage
\begin{widetext}
\begin{center}%
\textbf{{\large%
Supplemental Material
}}%
\end{center}%
\section{Estimating the BBR shift uncertainty}

We begin by writing the BBR Stark shift as
\begin{equation}
\nu_\mathrm{BBR}=-\frac{1}{2h\epsilon_0}\int\Delta\alpha(\omega)u(\omega)d\omega,
\end{equation}
where $h$ is Planck's constant, $\epsilon_0$ is the vacuum permittivity, $\Delta\alpha(\omega)$ is the differential (scalar) polarizability for the clock transition, and $u(\omega)$ is the spectral energy density of the thermal radiation at the location of the atoms. We note that tensor polarizabilities for the $^{171}$Yb clock states are identically zero, such that the atoms sample the local thermal radiation environment without directional bias. This legitimizes the formula above and justifies zero shift correction and uncertainty related to anisotropy of the thermal radiation environment (see Ref.~\cite{BelHinPhi14}), regardless of the extent to which the thermal radiation environment may deviate from an ideal BBR environment.

The spectral energy density of the thermal radiation is determined by the atoms' surroundings. We suppose the atoms are surrounded by a finite number of solid bodies, enumerated in the following by index $i$. The BBR Stark shift can be expressed as:
\begin{equation}
    \nu_\mathrm{BBR}=
    \left(\nu_\mathrm{BBR}\right)_0
    +\left(\delta\nu_\mathrm{BBR}\right)_{\delta T}
    +\left(\delta\nu_\mathrm{BBR}\right)_\mathrm{amb}
    +\left(\delta\nu_\mathrm{BBR}\right)_\mathrm{higher}.
    \label{Eq:nuBBRdeviations}
\end{equation}
The first term on the righ-hand-side of Eq.~(\ref{Eq:nuBBRdeviations}), $\left(\nu_\mathrm{BBR}\right)_0$, is the zeroth order approximation, corresponding to the BBR shift for the atoms in an ideal BBR environment with a chosen set temperature $T_\mathrm{set}$ of the shield. Uncertainty in this term is attributed entirely to the atomic response coefficients (i.e., $\nu_\mathrm{stat}$, $\nu_{\mathrm{dyn},6}$, \dots). The remaining terms represent corrections to this zeroth order approximation, and are discussed in the subsections below. In brief, the correction $\left(\delta\nu_\mathrm{BBR}\right)_{\delta T}$ accounts for temperature deviations relative to $T_\mathrm{set}$ for surfaces with direct line of sight to the atoms, while $\left(\delta\nu_\mathrm{BBR}\right)_\mathrm{amb}$ accounts for external thermal radiation leaking into the shield. $\left(\delta\nu_\mathrm{BBR}\right)_\mathrm{higher}$ is a ``higher-order'' correction that accounts for the combined effect of temperature deviations from $T_\mathrm{set}$ together with deviations of the line-of-sight surfaces from opaque, unit-emissivity surfaces. For these three correction terms, uncertainty due to atomic response is negligible. Uncertainties in these corrections are categorized as ``BBR environment'' uncertainty in Table~I of the main text. 

Our shield design aims to minimize reliance on thermal radiation simulations. Generally speaking, radiation simulations require numerous assumptions. An evaluation of the BBR shift that depends strongly on details of a radiation simulation, from physical assumptions (e.g., gray, diffuse surfaces) to input parameters (e.g., specific emissivity values), could be prone to errors. We emphasize that evaluation of $\left(\delta\nu_\mathrm{BBR}\right)_{\delta T}$ does not require radiation simulations, as only geometric solid angles for the line-of-sight surfaces and their temperatures are required. We do invoke radiation simulations, however, to assess conceivable upper limits for the magnitudes of $\left(\delta\nu_\mathrm{BBR}\right)_\mathrm{amb}$ and $\left(\delta\nu_\mathrm{BBR}\right)_\mathrm{higher}$, using a generous range of input parameters. We assign an uncertainty to each correction term equal to 100\% of the assessed upper limit. In essence, our design enables a conservative treatment of these corrections without real penalty. Regardless of $T_\mathrm{set}$, the uncertainties in $\left(\delta\nu_\mathrm{BBR}\right)_\mathrm{amb}$ and $\left(\delta\nu_\mathrm{BBR}\right)_\mathrm{higher}$ are overshadowed by uncertainty in $\left(\delta\nu_\mathrm{BBR}\right)_{\delta T}$.

\subsection{Temperature deviations, $\left(\delta\nu_\mathrm{BBR}\right)_{\delta T}$}
As mentioned earlier, the correction $\left(\delta\nu_\mathrm{BBR}\right)_{\delta T}$ accounts for temperature deviations relative to $T_\mathrm{set}$, exclusively for surfaces with direct line of sight to the atoms. For the purposes of this correction, the direct-line-of-sight surfaces are treated as opaque, blackbody surfaces. Uncertainty in this correction is attributed to temperature gradients across the shield components, as well as measurement uncertainty associated with the RTD readings. Given small temperature deviations relative to $T_\mathrm{set}$, this correction can be well approximated by
\begin{equation}
\left(\delta\nu_\mathrm{BBR}\right)_{\delta T}
\approx
\nu_\mathrm{stat}\left(\frac{4T_\mathrm{set}^3}{T_0^4}\right)\delta T = \nu_\mathrm{stat}\left(\frac{4T_\mathrm{set}^3}{T_0^4}\right) \frac{1}{4\pi}
\sum_i\Omega_i\delta T_i,
\label{Eq:dnuBBRdT}
\end{equation}
where $\Omega_i$ is the geometric solid angle subtended by the $i$th body (with temperature $T_i$) from the vantage point of the atoms, $\delta T_i=T_i-T_\mathrm{set}$, and $T_0=300$~K. This result is obtained by taking the static approximation for the differential polarizability, $\Delta\alpha(\omega)\rightarrow\Delta\alpha(0)$, together with $T_i^4-T_\mathrm{set}^4\approx4T_\mathrm{set}^3\delta T_i$. The cubic temperature dependence seen here ($\propto T_\mathrm{set}^3$) is one of the main motivations for pursuing cryogenic clock operation.

\renewcommand{\arraystretch}{1.5}
\setlength{\tabcolsep}{5pt}

\begin{table}
    \centering
    \begin{tabular}{l c c c}
        \toprule
        $i$ & $\Omega_i/4\pi$ & $\quad[\delta T_{i,\mathrm{min}}, \delta T_{i,\mathrm{max}}]$ & $\quad [\left(\delta\nu_\mathrm{BBR}\right)_{\delta T,\mathrm{min}},\left(\delta\nu_\mathrm{BBR}\right)_{\delta T,\mathrm{max}}]\quad$\\
        \midrule
        Shutter sphere   & $0.7648$ & $[11, 71]$    &$[-2.96,-0.46]$\\
        Shield halves    & $0.1181$ & $[51, 248]$   &$[-1.60,-0.33]$\\
        Electrodes       & $0.1070$ & $[96, 303]$   &$[-1.77,-0.56]$\\
        Lattice window   & $0.0101$ & $[210, 1100]$ &$[-0.61,-0.12]$\\
        \midrule
        Sum/Weighted sum & $1.0000$ & $[27, 127]$   &$[-6.93,-1.47]$\\
        Mean (Standard deviation)   & & $77 (29)$ & $-4.2 (16)$\\
        \bottomrule
    \end{tabular}
    \caption{Fractional geometric solid angles for each of the shield parts that has direct line of sight to the atoms and corresponding temperature limits, in mK, retrieved from the experiment-informed thermal simulations at $T_\mathrm{set}=77$~K. The range of the electrodes is shifted by 50~mK higher and expanded by 10~mK to account for the observed variation in the dedicated temporary setup. The last column shows the limits of $\left(\delta\nu_\mathrm{BBR}\right)_{\delta T}$, in units of $10^{-21} \nu_0$ where $\nu_0$ is the unperturbed clock frequency. The limits are weighted by the fractional solid angle for each shield part. The bottom right cell shows the limits of the $\left(\delta\nu_\mathrm{BBR}\right)_{\delta T}$ uniform probability distribution.}
    \label{Tab:SA and temp limits}
\end{table}

RTD measurement uncertainty sources are detailed in Sec.~\ref{sec: other uncertainty sources}. Since they are uncorrelated and not specific to any specific body of the shield, they are added in quadrature to the other uncertainty sources.

In the closed configuration, the atoms are surrounded by the three main copper body parts (the shutter sphere and the two shield halves) and the small optical access fitted with cryogenic N-BK7 substrates for the lattice and clock spectroscopy beams. These body parts subtend the entire geometric solid angle around the atoms, and the fractional solid angle subtended by each is presented in Table \ref{Tab:SA and temp limits}. At $T_\mathrm{set}=77$~K, we measure a temperature difference of 55 mK across the shutter sphere and $\approx$ 85~mK across the thermal extremities of both the stationary shield halves. The electrode temperatures were measured previously in a dedicated temporary setup and were found to be $\approx$50~mK higher than the mean temperature of the shield half it is mounted to. The RTD mounted inside the lattice window measures $\approx$130~mK above the mean shield temperature.

We refer to thermal simulation to confirm (and inform if necessary) that the RTDs faithfully represent the temperature extremes of the internal surfaces of the shield. Initial simulations in the design phase relied on rough estimates of the thermal resistance of shield supports inside the vacuum chamber and polished copper emissivities. The measured temperatures on the shield parts agree within a factor of 2 with these initial simulations. Better agreement is obtained when the simulations are informed with the measured thermal loads, derived in Sec.~\ref{sec: Thermal_load}. The agreement is particularly good at the 5\% level for the stationary radiation shield since it is subjected to the vast majority ($>$95\%) of the total thermal load. For the shutter sphere, the thermal load is small ($\sim$100~mW) and thus is harder to estimate from the calculations in Sec.~\ref{sec: Thermal_load}. However, most of the sphere body is enclosed in the shield, and thus the thermal load is almost entirely applied to its exposed drive tip. We find that by simulating a 110~mW of thermal load at the tip reproduces the measured thermal differences of the RTDs embedded in the shutter sphere. 

The experiment-informed thermal simulations confirm that the RTDs faithfully represent the temperature extremes of the shutter sphere surfaces with direct line of sight to the atoms within 10\%. For the stationary shield, the internal surfaces (including those of the electrodes) exhibit a $2\times$ larger temperature extremes than seen by the embedded RTDs. Intense lattice light can cause localized heating due to absorption in the window substrate \cite{SCHOTT}. We simulate the effect by modeling the beam as a cylindrical heat source with the lattice waist diameter centered at the lattice incident location onto the window substrate, and fix the temperature of the substrate contact to the shield temperature. In the calculations described below, we use the larger temperature extremes observed in these experiment-informed thermal simulations.

To calculate the uncertainty on the correction $\left(\delta\nu_\mathrm{BBR}\right)_{\delta T}$, we first estimate its probability distribution. For each body, we assume a uniform distribution for $\delta T_i$ between $\delta T_{i,\mathrm{min}}$ and $\delta T_{i,\mathrm{max}}$, denoting the minimum and maximum temperature deviations from $T_\mathrm{set}$, respectively. These limits are obtained from the experiment-informed thermal simulations described above, and are presented in Table~\ref{Tab:SA and temp limits}. We emphasize here that in deriving the probability distribution of $\left(\delta\nu_\mathrm{BBR}\right)_{\delta T}$, we do not combine uncorrelated contributions to the correction from each body. Instead, we impose perfect correlation between the contributions from each, meaning all bodies must simultaneously sample either the minimum, maximum, or the same relative position (or interpolation point) between the two temperature extremes. This ensures that in the limit when all parts have the same temperature extremes, the resulting distribution matches the initial uniform distribution, thus avoiding the non-physical reduction of probability distribution spread by an arbitrary division of the surfaces into smaller subsets. Further, if a part has zero uncertainty in its temperature, it does not contribute to the spread of the correction, and if a part has a small solid angle, it has small impact on the overall distribution. With this perfect correlation, which we regard as a worst case scenario, it is trivial to prove that the resulting distribution of $\left(\delta\nu_\mathrm{BBR}\right)_{\delta T}$ is also uniform in a range bounded by the two values $\left(\delta\nu_\mathrm{BBR}\right)_{\delta T,\mathrm{max}}=-1.47\times 10^{-20}$ obtained by substituting $\delta T_i \rightarrow\delta T_{i,\mathrm{min}}$ and  $\left(\delta\nu_\mathrm{BBR}\right)_{\delta T,\mathrm{min}}=-6.93\times 10^{-20}$ obtained by substituting $\delta T_i \rightarrow\delta T_{i,\mathrm{max}}$ in Eq.~(\ref{Eq:dnuBBRdT}). The uncertainty for the correction term $\left(\delta\nu_\mathrm{BBR}\right)_{\delta T}$ is taken as the standard deviation of the uniform distribution, or $1/\sqrt{12}$ of the interval. This corresponds to a fractional frequency uncertainty of $1.6\ \times 10^{-20}$, which is equivalent to an uncertainty of 29~mK on the mean effective temperature at the atoms’ location.

\subsection{The residual external thermal radiation leak, or $\left(\delta\nu_\mathrm{BBR}\right)_\mathrm{amb}$}
\label{sec:leak}
During spectroscopy, the shield is in the closed position. Although the atoms don’t have direct exposure to the external thermal environment, a residual fraction of external thermal radiation can reach the atoms by transmission through the cryogenic optical substrate or multiple reflections on the coated internal nested structure of the shield. In this section, we estimate an upper bound on the Stark shift that this residual radiation causes. The treatment below pertains to operation at 77~K but is also repeated to estimate the effect at 298.5~K and 318.5~K. In brief, we estimate the effective solid angles of various apertures $\Omega_\mathrm{eff}^{a}$ to the external environment through a reverse ray-tracing Monte Carlo simulation and assume a conservative range of emissivity values of the thermal coatings and transmission data of the window substrates over all relevant wavelengths. We use the results to place an upper bound on the shift resulting from the leaking external BBR, as detailed below.

\begin{figure*}
\centering
\includegraphics[width=0.7\textwidth]{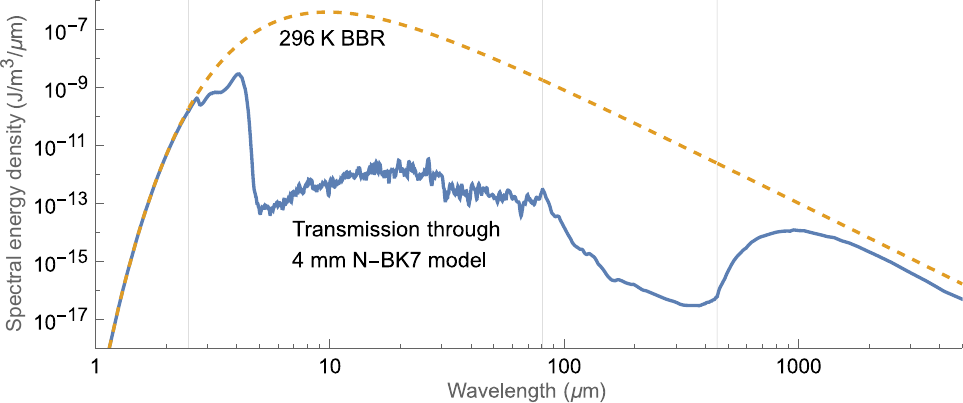}
\caption {BBR spectrum at room temperature of $T_\mathrm{ext}=296$ K and the transmitted fraction through a 4 mm N-BK7 substrate described by the transmission model $t(\lambda)$. For wavelengths ranging from $1.06\ \mathrm{\mu m}$ to $2.5\ \mathrm{\mu m}$, we utilize transmission data provided by the manufacturer \cite{SCHOTT}. For wavelengths between $2.5\ \mathrm{\mu m}$ and $81\ \mathrm{\mu m}$, we conservatively adopt the instrument-noise-limited transmission data for the 3.2 mm thick substrate (scaled to 4~mm) from Ref.~\cite{al2016strontium} as an upper bound, similarly for $81\ \mathrm{\mu m}<\lambda\leq 450\ \mathrm{\mu m}$ from \cite{islam2020experimental}.  For the range from $450\ \mathrm{\mu m}$ to $5000\ \mathrm{\mu m}$, the transmission data from Ref.~\cite{naftaly2007terahertz} are used. For all other wavelengths, we assume full transmission. Vertical lines mark the boundaries between different literature transmission sources.}
\label{fig:Transmission}
\end{figure*}

The reverse ray tracing starts with a sufficiently large number of $N$ rays at the atoms, emitted uniformly in all directions from the location of the atoms at the center of the shield. The rays propagate outward and undergo multiple reflections on the internal surfaces of the shield, which are assumed to be Lambertian (i.e. diffuse). For each ray, the number of reflections $n$ is tracked. The rays propagate until they meet one of three possible terminations: (1) ``captured" by the windows on any of the three MOT/lattice apertures, (2) escape the shield through the single aperture pair for the atomic beam or (3) disappear at the $62^\mathrm{nd}$ reflection. The third condition ensures that the fractional shift caused by the radiation carried by a single ray is negligible at the low $10^{-21}$ level. The number of reflections corresponds to 6 orders of magnitude reduction of radiation power when the ray reflects off surfaces with the lowest emissivity considered below ($\epsilon=0.2)$. The simulation ends when all emitted rays have terminated. Given the value of black coating emissivity $\epsilon$, we use the statistics of rays terminated in (1) and (2) to estimate the effective solid angles of each respective apertures $\Omega_\mathrm{eff}^{a} $ according to:
\begin{equation}
    \frac{\Omega_\mathrm{eff}^{a} (\lambda)}{4\pi}=\sum_n\frac{C_n^a  [1-\epsilon(\lambda)]^n}{N},
\end{equation}
where $C_n^a$ is the number of rays that underwent exactly $n$ reflections and terminated on aperture $a$, and $[1-\epsilon(\lambda)]^n$ is the power attenuation factor associated with $n$ reflections. Given the values of $\epsilon(\lambda)$ reported in Ref.~\cite{adibekyan2017high}, we conservatively set $\epsilon(\lambda\leq100 \ \mu $m$)=0.8$. There is little to no emissivity data available for the black coating at larger wavelengths, and thus we assume a conservative $\epsilon(\lambda > 100\ \mu $m$)=0.2$.

We now shift from the reverse ray tracing perspective---which considers radiation paths traced backward-in-time from the internal volume---we to the forward-in-time perspective. For evaluation purposes, we classify the shield apertures based on the type of suppression that the external thermal radiation experiences before reaching the internal volume of the shield. The type of suppression can be by absorption by the N-BK7 windows or multiple reflections off the coated surfaces of the shield. External thermal radiation through a given aperture is suppressed by one or both modes. We identify three classes of apertures: (1) apertures for the atomic beam, fitted with high-aspect-ratio coated tubes, (2) apertures for the MOT beams, and (3) apertures for lattice and clock beams. External thermal radiation passing through the apertures in class (1) and (2) is extensively suppressed by multiple reflections off the internal coated surfaces of the shield. For apertures in classes (2) and (3), external thermal radiation is absorbed into one or two of the 4 mm-thick N-BK7 substrates. These substrates are thermally coupled to the shield via pressed indium to maintain their temperatures at or close to the shield temperature. Specifically, apertures of class (3) are fitted with two stacked 4 mm substrates that are independently coupled to the shield, whereas apertures in class (2) are fitted with a single 4 mm substrate each. We note that the vertical MOT optical access is fitted with two substrates and thus possesses better absorption to the penetrating external radiation than the horizontal MOT apertures, but we still classify them as class (2) with a single substrate to simplify the calculations. This simplification only inflates the contribution from that source, and only slightly increases the upper bound estimated in this treatment. 

To evaluate the performance of N-BK7 substrates in blocking external thermal radiation, we construct an aggregate transmission data of N-BK7 material guided by the literature \cite{SCHOTT,al2016strontium,islam2020experimental,naftaly2007terahertz}. The references cited cover the range from $1.06\ \mathrm{\mu m}$ up to $5000\ \mathrm{\mu m}$. A given reference might report values over a wavelength range that overlaps with another reference, and some only provide upper bounds on the transmission. We found that the references agree fairly well in the overlapping wavelength ranges. We construct a conservative aggregate $t(\lambda)$ by assuming the more transmissive value or the upper bound in the overlapping wavelength ranges and full transmission where the substrate is mostly transmissive or no data is available, covering the entire electromagnetic spectrum. The room-temperature BBR spectrum transmitted through a single 4 mm-thick N-BK7 described by the model transmission $t(\lambda)$ is shown in the Fig. \ref{fig:Transmission}. 

The fractional Stark shift due to the radiation leaking to the atoms through the aperture $a$ is

\begin{equation}
    \left(\delta\nu_\mathrm{BBR}\right)_{\mathrm{amb},i}=-\int_{0}^{\infty} \frac{\Omega_\mathrm{eff}^{a} (\lambda)}{8\pi\epsilon_0 h} \Delta\alpha_\lambda(\lambda)\  t(\lambda)^i\ [u_\lambda(\lambda,T_\mathrm{ext})-u_\lambda(\lambda,T_\mathrm{set})]\ d\lambda,
    \label{eq: BBR leak shift}
\end{equation}
where $\Delta\alpha_\lambda(\lambda)$ is the clock differential polarizability, $u_\lambda(\lambda,T)$ is the spectral energy density of an ideal BBR at temperature $T$, and $i$ is the number of 4 mm substrates the external radiation passes through. $\Delta\alpha_\lambda(\lambda)$ relates to its angular frequency forms $\Delta\alpha(\omega)$ by a simple spectral variable substitution. In Eq.~(\ref{eq: BBR leak shift}), the spectral energy density $ u_\lambda(\lambda,T_\mathrm{set})$ of the BBR environment created by the shield is subtracted to account for the radiation escaping the shield that would otherwise contribute to the whole BBR shift. Such accounting becomes important when estimating the bounds on the residual external radiation shift at non-cryogenic temperatures of the shield. For example, operating the shield close to room temperature results in a balance between the escaping BBR and the external radiation going into the shield. In other words, the perturbation to the BBR shift due to the leaking external thermal radiation through the apertures when the shield is close to room temperature is smaller than when the shield is far from room temperature. Such balance is the basis that room-temperature radiation shields rely on to achieve low uncertainty despite possessing much larger exposure to the external environment compared to cryogenic BBR solutions. Although this large exposure to the external thermal environment is typically associated with a large uncertainty in estimating $\Omega_\mathrm{eff}^a$, the uncertainty gets suppressed by the smaller factor $\left[u_\lambda(\lambda,T_\mathrm{ext})-u_\lambda(\lambda,T_\mathrm{shield})\right]$ in Eq.~(\ref{eq: BBR leak shift}) when $T_\mathrm{shield}\approx T_\mathrm{ext}$. When the shield is operated at higher than room temperature, such as at 318.2~K, the correction $(\delta\nu/\nu_0)_i$ flips sign. It is interesting to note that in such a case the uncertainty is dominated by the lack of knowledge on how much BBR escapes the shield (as a result of uncertainty on $\Omega_\mathrm{eff}^a$), and not by how much external thermal radiation has leaked into the shield. We assign the absolute values of the shift at all three temperatures as 100\% uncertainty.

Table \ref{tab:Aperture Classes} details the Stark shift induced by leaking external radiation through each aperture class. If the radiation is subject to reflections off black coatings after it passes through a given aperture class, $\Omega_\mathrm{eff}^{a} (\lambda)$ is a step function with two different constant values before and after $\lambda = 100\ \mathrm{\mu m}$, a direct consequence of the two constant values $\epsilon(\lambda\leq100 \ \mu $m$)=0.8$ and $\epsilon(\lambda\geq100 \ \mu $m$)=0.2$ assumed. This is the case for atomic beam and MOT apertures. For the small lattice/clock apertures, $\Omega_\mathrm{eff}^{a} (\lambda)$ reduces to the plain geometric solid angle  and is independent of the wavelength to first approximation, since the external radiation passing through the lattice/clock apertures is suppressed mostly due to absorption by the two stacked windows. We confirm with ray tracing that the variation of the solid angle subtended by the lattice aperture's due to a displaced atoms' position or reflections off the black coatings is bounded to $<$10\%. We thus assume a solid angle inflated by 10\% to account for such effects. 

\renewcommand{\arraystretch}{1.5}
\setlength{\tabcolsep}{5pt}

\begin{table}[h!]
\centering
\begin{tabular}{|c|c|c|c|c|c|c|}
    \hline
    \multirow{2}{*}{Aperture Class} & \multicolumn{2}{c|}{$\Omega_\mathrm{eff}^a(\lambda)/4\pi$} & \multirow{2}{*}{\makecell{No. of N-BK7\\ substrates ($i$)}} & \multicolumn{3}{c|}{$\left(\delta\nu_\mathrm{BBR}\right)_{\mathrm{amb},i}/ \nu_0$} \\
       & $\lambda < 100\ \mathrm{\mu m}$ & $\lambda > 100\ \mathrm{\mu m}$ & & $\lambda < 100\ \mathrm{\mu m}$ & $\lambda > 100\ \mathrm{\mu m}$ & all $\lambda$ \\
    \hline
    \multirow{1}{*}{Atomic beam} & $3.1\times 10^{-8}$ & $4.0\times 10^{-5}$ & 0 & $-7.2\times 10^{-23}$ & $-3.8\times 10^{-22}$ & $-4.5\times 10^{-22}$ \\
                            MOT  & $1.2\times 10^{-3}$ & $2.8\times 10^{-2}$ & 1 & $-1.2\times 10^{-21}$ & $-1.0\times 10^{-22}$ & $-1.3\times 10^{-21}$ \\
    Lattice/Clock                & \multicolumn{2}{c|}{$1.1\times 10^{-2}$}  & 2 & \multicolumn{2}{c|}{$-1.5\times 10^{-21}$}   & $-1.5\times 10^{-21}$ \\
    \hline
\end{tabular}
\caption{Estimates of the external thermal radiation leak for each of the three classes of apertures on the cryoshield at $T_\mathrm{set}=77$~K, and the associated shifts in fractional clock frequency.}
\label{tab:Aperture Classes}
\end{table}

Summing the contributions from all apertures, we place an upper bound of $3.3\times10^{-21}$ on the absolute value of the fractional shift due to the residual external thermal radiation during spectroscopy, and assume the whole shift as its uncertainty. For reference, we note that setting a more conservative $\epsilon(\lambda > 100\ \mu $m$)=0.1$ inflates this upper bound to $6.5\times10^{-21}$.

\subsection{Higher order corrections, $\left(\delta\nu_\mathrm{BBR}\right)_\mathrm{higher}$ }
As mentioned earlier, the $\left(\delta\nu_\mathrm{BBR}\right)_\mathrm{higher}$ is a ``higher-order'' correction that accounts for the combined effect of (a) deviation of direct-line-of-sight surfaces from being opaque, blackbody surfaces and (b) temperature deviations of the various shield components from $T_\mathrm{set}$. This correction captures the residual shift that violate the assumptions involved in evaluating $\left(\delta\nu_\mathrm{BBR}\right)_{\delta T}$ and $\left(\delta\nu_\mathrm{BBR}\right)_\mathrm{amb}$. It also captures shifts from bodies that don't have direct line of sight to atoms whose temperatures are far from both $T_\mathrm{ext}$ and $T_\mathrm{set}$.

There are a few sources that contribute to this term. First, deviations from the geometric solid angles of the internal surfaces in Eq.~(\ref{Eq:dnuBBRdT}) due to a possible displacement of the atoms' position from the shield center or the non-unit emissivity of the black coating or the window substrate. Ray tracing analysis confirms that the subtended solid angles of the internal surfaces remain largely insensitive to both the atoms' position and the emissivities of internal surfaces. Specifically, for any emissivity value $\epsilon\in[0.8,1]$, whether for the black coatings or windows, or atomic position displacements up to 5 mm from the geometric center of the shield, absolute variations in the fractional solid angle remain below 0.5\%. The fractional shift from both effects are bounded to $<2\times10^{-21}$ at $T_\mathrm{set}=77$~K. Negligible shift bounds are also established for the other shield set temperatures.

The cryogenic windows that directly absorb the ambient radiation develop thermal gradients across their internal surfaces, which emit thermal radiation at slightly different temperature that has the potential to perturb the BBR environment at the atoms' location. We emphasize that none of those windows has direct line of sight to the atoms. The first class of these are the the single 4 mm substrate on the horizontal MOT apertures, which has indirect exposure after multiple reflections on the black coatings. To estimate the potential shift, we simulate the thermal gradients of the cold window subject to the radiative load due to absorption of the ambient thermal radiation and estimate that the windows develop a mean temperature difference $(\overline{T_\mathrm{win}}-T_\mathrm{shield})$ on its inner surface of $1.9$~K, $-0.06$~K, and $-0.6$~K at $T_\mathrm{set}$=77~K, 298.5~K and, 318.2~K, respectively. The fractional shift induced is $-3.3\times10^{-21}$, $2.4\times10^{-21}$ and $2.9\times10^{-20}$ respectively. The second class are the outer windows on the doubly-stacked arrangement on the vertical direction. Radiation from these can transmit through the inner window and reach the atoms. The induced shift is well below $10^{-23}$ for up to 10~K of temperature increase of the outer window at $T_\mathrm{set}=77$~K. Negligible shifts are induced at other $T_\mathrm{set}$ as well.

\subsection{Other uncertainty sources}
\label{sec: other uncertainty sources}

\subsubsection{RTD calibration and its stability against thermal shocks}
The negative-temperature-coefficient thin-film RTDs are calibrated by the vendor \cite{commercialSM}, with uncertainties provided at each temperature. All the RTDs used on the shield (except for the sensor embedded in the optical substrate) were thermally shocked at least 30 times by repeated immersion in liquid nitrogen followed by warm-up to room temperature. The thermal shocks were performed to verify calibration stability of the thin-film RTDs and were motivated by significant calibration shifts observed earlier upon thermal cycling of platinum wire-wound RTDs. After the thermal shocks and before the RTD installation in the shield, we verify the agreement between the sensors by embedding them in a single copper block that is later immersed in liquid nitrogen bath and then in a water-ice mixture. The sensors exhibit agreement at better than the 1$\sigma$ level of the vendor specifications of 6~mK at 77~K and 18~mK at 273~K. The temperatures were read out using the same unit later used on the experiment for temperature control. Further, the sensors' absolute calibration was also verified against the ice point of water at the 2$\sigma$ level.

\subsubsection{Electronic measurement error}

The electronic measurement error of the 4-wire multimeter is improved beyond the manufacturer specifications by regularly calibrating each of the instrument's eight inputs against a standard decade resistor box \cite{commercialSM}. We perform the calibration at resistances above and below the span of resistances of the RTDs at 77~K, 298~K, and 318~K. We then use linear interpolation to correct the instrument temperature readings in post-processing. We find that the manufacturer uncertainty specifications for the 4-wire multimeter is highly conservative. At 298~K and 318~K, it can be safely reduced by $> 12\times$ using the method described. The uncertainty quoted in the main text is the combined uncertainty of the standard resistor and transfer uncertainty estimated by the negligible drift rate of the instrument measured over the course of 3 months. At 77~K, the uncertainty quoted is more conservative and is assigned the largest observed \textit{shift} on any of the 8 inputs of the device. This is still $\sim 7\times$ smaller than the manufacturer specifications and larger than the uncertainty of the standard resistor. 

\subsubsection{Self-heating}
The RTDs embedded in the shield require a constant excitation current to measure their resistance and infer the temperature. However, current passing through the detector produces unwanted ohmic heating that results in a bias in the temperature measurement. This effect is known as self-heating and could potentially cause biases larger than the sensor calibration uncertainties. For thin film RTDs such as the ones used in this work, the sensing element is deposited directly on a sapphire base attached to the alumina sensor body. Such construction provides good thermal contact of the detector element to the sensor capsule, even in vacuum, largely mitigating the self-heating effect. This is in contrast with the strain-free mounting needed for bulk and thick wire sensing elements such as platinum RTDs, which suffer from limited thermal interface between the sensing element and sensor body, especially in vacuum. Care must also be exercised to ensure sufficient thermal contact between the sensor capsule and the measured sample. For all sensors on the shield, we immerse the sensor body in a hole drilled into the shield component and filled with perfluoropolyether (PFPE) based grease. This mounting ensures efficient and reversible thermal attachment to the shield component. The 4-wire multimeter excitation current automatically switches between 3 $\mu$A at 77~K and 30 $\mu$A at 298~K and 318~K to balance between reducing the self-heating effect and improving the signal-to-noise ratio of the readout voltage. We estimate the self-heating uncertainty by driving the RTD  with a much higher 1 mA excitation current and measuring the apparent rise in the sensor temperature. Such measurement provides a lever arm anywhere between $10^3$ and $10^5$ to estimate the self-heating at the operating current. The whole self-heating effect is negligible at all temperatures and is taken as 100\% uncertainty.

\subsubsection{Parasitic conduction and radiation loads on temperature sensors}
Thermal anchoring of temperature sensor wire is crucial to ensure that thermal load from the wire doesn't cause unwanted shifts to the temperature measurements. One recommendation is to use $\sim$ 1.1 cm temper length \cite{ekin2006experimental}, however these estimations neglect the thermal radiation load on the wire since it assumes that the entire wire length ($\sim$25 cm) extends in a cryogenic environment, with negligible radiative load. This is not the case in our setup where the wire is subject to ambient radiation, so a re-evaluation of temper length is required to account for the much larger radiative load. We run thermal simulations on a cylindrical shaped wire, and we find that including the radiative load (emissivity of polyimide wire insulation is assumed to be unity) increases the thermal leak at the cryogenic end of the wire by almost 6 times compared to conductive load only. Referring to the original calculations in Ref.~\cite{hust1970thermal} used to derive the recommended temper lengths, we calculate that such extra radiative load requires increasing the temper length by only 30\%. We note that the shortest temper length for two of the eight sensors we have is more than 2 cm. Additionally we use the much more thermally conductive indium pressed against the wire as our thermal ``adhesive," compared to varnish or epoxy assumed in Refs.~\cite{hust1970thermal,ekin2006experimental}. We also run the calculations to confirm that this effect is insignificant at 298.5K and 318.2K.

Finally, care should be exercised to protect exposed sensor wire from thermal radiation even after the conductive thermal tempering. We found that even 1 mm of exposed wire on the stationary shield sensors can lead to measurable temperature reading shifts at cryogenic temperatures. To prevent that, we anchor the entire length of the wire with pressed indium up to the sensor, and extend pieces of thermally anchored indium sheets to cover the holes into which the sensors are embedded, while making vents in the indium sheets to prevent trapped air volumes in vacuum. During the assembly of the shield, we verify the efficacy of the setup by hovering a hot emissive object (a glass beaker with boiling water) locally at each sensor and observing no appreciable increase in temperature beyond the overall increase in the bulk temperature of the copper piece as witnessed by the other sensors on the shield.

\subsubsection{BBR Zeeman shift}

In addition to a Stark shift, BBR also induces a Zeeman shift. For the Zeeman shift, the dominant contribution is from the (magnetic dipole) coupling between the $6s6p\,{^3P_0}$ clock state and the neighboring $6s6p\,^3P_1$ state. At room temperature, the characteristic BBR frequency ($\sim k_BT/h$) is comparable to the fine structure splitting between these states. Consequently, calculating this contribution requires evaluation of the Farley-Wing function~\cite{FarWin81,PorDer06}, rather than a truncated version of its asymptotic expansion. We also consider contributions beyond the $6s6p\,{^3P_0}$ -- $6s6p\,{^3P_1}$ contribution, inclusive of the diamagnetic contribution. These ``other'' contributions are satisfactorily captured by the leading asymptotic term of the Farley-Wing function (amounting to a ``static'' approximation, with $T^4$ dependence). To evaluate the different contributions, we use values tabulated for the magnetic dipole polarizabilities of the clock states in Ref.~\cite{Tobyaccepted} (see online supplemental material). Results are presented in Table~\ref{Tab:BBRZeeman} for the three temperatures considered in the main text: 77.0~K, 298.5~K, and 318.2~K. Each value in the table is accurate to better than 1\%, implying BBR Zeeman shift uncertainty $<\!10^{-23}$ at 77.0~K and $<\!10^{-21}$ at 298.5~K and 318.2~K. BBR shifts due to higher multipolar couplings (e.g., electric quadrupole) are neglected~\cite{PorDer06}.

Finally, we note that the BBR Zeeman shift is typically neglected altogether in OLCs, being overshadowed by uncertainties in the BBR Stark shift. Interestingly, we see that the room-temperature BBR Zeeman shift for Yb ($\approx-3\times10^{-20}$) exceeds the total BBR shift uncertainty for our Yb OLC when the shield is operated at cryogenic temperatures ($\approx2\times10^{-20}$).

\begin{table}[t]
\caption{BBR Zeeman shift for the Yb clock transition evaluated at the temperatures used in the main text (clock frequency $\times10^{-21}$). The contribution from the $6s6p\,{^3P_0}$ -- $6s6p\,{^3P_1}$ coupling is given separately from other contributions.}
\label{Tab:BBRZeeman}
\begin{ruledtabular}
\begin{tabular}{lccc}
contribution	& 77.0~K	& 298.5~K	& 318.2~K	\\
\hline\vspace{-2.5mm}\\
$6s6p\,{^3P_0}$ -- $6s6p\,{^3P_1}$
				& $-0.455$	& $-31.7$	& $-28.9$
\\
other			& $0.00651$	& $1.47$	& $1.90$
\end{tabular}
\end{ruledtabular}
\end{table}

\section{Faraday shielding and constraining the stray-field dc Stark shift}

As noted in the main text, the BBR shield doubles as a Faraday shield. Thus, the atoms are not only furnished with a near-ideal BBR environment, but also one that is free of dc electric fields.  The shield assembly includes six circular electrodes (labeled as item 3 in Fig.~1b of the main text), which are grounded in normal operation to form part of the Faraday shield. The electrodes are aligned along three mutually orthogonal axes, with the axes intersecting at the center of the shield. There are two electrodes per axis, positioned on opposite sides of the shield. Each electrode is affixed to one of the two outer stationary shield bodies; when the shutter sphere is in the closed (spectroscopy) position, apertures on the shutter sphere align with the electrodes. Voltages can be applied to the electrodes with an external voltage source, with auxiliary components (e.g., vacuum feedthroughs, electrode sapphire insulating spacers, electrical leads) accommodating applied voltages as large as $\pm2$~kV. The two outer stationary shield bodies and the shutter sphere are permanently grounded.

The electrodes provide a means to assess the efficacy of the Faraday shielding, through measurement of a dc Stark shift of the clock transition \cite{lodewyck2012observation}. When equal but opposite voltages are applied to the electrodes of a given axis, the result is an applied field at the center of the shield that nominally aligns with the axis. Given asymmetry in the shield geometry, applied fields associated with the three axes are not expected to be exactly orthogonal to one another. We follow the dc Stark analysis outlined in Ref.~\cite{BelZhaMcG18} (see main text and online supplemental material), which does not presuppose orthogonality of these applied fields.

Table~\ref{Tab:dcStark} presents dc Stark shift data for various configurations of the applied voltages. In each case, the clock frequency is measured relative to the fully grounded arrangement. The voltage configuration is specified by a list containing three entries, with each entry identifying one of the three axes. For the first entry (i.e., axis), $+$ indicates that $+2$~kV is applied to one electrode and $-2$~kV is applied to the opposing electrode, $-$ indicates the same but with opposite polarity, and $0$ indicates that both electrodes are grounded. This labeling scheme holds for the other axes as well, except that applied voltages of $\pm250$~V replace $\pm2$~kV for the third axis. A minimum of nine configurations are required to fully evaluate the stray-field dc Stark shift. We employ eighteen configurations, with nine of the configurations being related to the other nine by polarity reversal. While the ``extra'' polarity-reversed configurations are not strictly needed, their inclusion affords a useful rudimentary analysis of the stray-field dc Stark shift. Namely, an observed frequency difference between opposite-polarity conditions immediately indicates the presence of a stray dc electric field. Inspection of our data reveals no statistically significant frequency difference between polarity-reversed configurations, indicating effective Faraday shielding. For a more quantitative analysis, we use a Monte Carlo procedure to map measurement uncertainty into a probability distribution for the stray-field dc Stark shift, as described in Ref.~\cite{BelZhaMcG18}. Neglecting any effects associated with electric field gradients across the atomic ensemble (discussed below), we find a 68.3\% confidence interval of $[-5,0]\times10^{-20}$ and a 95.5\% confidence interval of $[-11,0]\times10^{-20}$ for the stray-field dc Stark shift, in units of the clock frequency. The distribution is visually similar to the distribution presented in Ref.~\cite{BelZhaMcG18}, with the shift effectively being constrained to negative values.

\newcommand{\e}[1]{\times10^{#1}}
\newcommand{\tbl}{5mm}
\newcommand{\tbox}[3]{(\makebox[\tbl][c]{#1},\makebox[\tbl][c]{#2},\makebox[\tbl][c]{#3})}
\begin{table}[t]
\caption{Frequency shifts induced by applying voltages to the electrodes, measured relative to the fully grounded configuration. See the text for voltage specifications corresponding to different configuration labels. Left and right data columns correspond to upper and lower signs in the configuration and represent opposite polarity conditions.}
\label{Tab:dcStark}
\begin{ruledtabular}
\begin{tabular}{lcc}
configuration		& \multicolumn{2}{c}{induced shift $\left(\text{clock frequency}\e{-16}\right)$}		\\
\hline\vspace{-2.5mm}\\
\tbox{$\pm$}{0}{0}				& $-2.83(11)$		& $-2.94(11)$	\\
\tbox{0}{$\pm$}{0}				& $-3.10(11)$		& $-2.94(12)$	\\
\tbox{0}{0}{$\pm$}				& $-2.80(14)$		& $-2.85(11)$	\\
\tbox{$\pm$}{$\pm$}{0}			& $-6.21(13)$		& $-6.21(11)$	\\
\tbox{$\pm$}{$\mp$}{0}			& $-5.57(9)$		& $-5.70(12)$	\\
\tbox{$\pm$}{0}{$\pm$}			& $-5.80(11)$		& $-5.93(7)$	\\
\tbox{$\pm$}{0}{$\mp$}			& $-5.78(12)$		& $-5.68(10)$	\\
\tbox{0}{$\pm$}{$\pm$}			& $-6.03(11)$		& $-5.99(8)$	\\
\tbox{0}{$\pm$}{$\mp$}			& $-5.83(10)$		& $-5.86(7)$	
\end{tabular}
\end{ruledtabular}
\end{table}

The Monte Carlo technique simultaneously allows evaluation of the angles between the applied fields~\cite{BelZhaMcG18}. We find $\cos\theta_{12}=-0.049(9)$, $\cos\theta_{13}=-0.014(8)$, and $\cos\theta_{23}=-0.013(8)$, where $\theta_{ij}$ is the angle between applied fields associated with the $i$-th and $j$-th axes. This implies that the applied fields are within a few degrees of being orthogonal to one another.

As discussed in Ref.~\cite{BelZhaMcG18}, error can enter the stray-field dc Stark shift analysis due to electric field gradients (stray or applied) coupled with the finite spatial extent of the atomic ensemble.  Generally, this is more of a concern for systems that exhibit an appreciable nonzero stray-field dc Stark shift. Our tight constraint on the shift with respect to zero allows a conservative treatment of this effect. Following Ref.~\cite{BelZhaMcG18}, we multiply the aforementioned distribution by $(1+\eta)$, taking $\eta$ to have a uniform distribution between zero and unity. For the resulting distribution, we find a 68.3\% confidence interval of $[-7,0]\times10^{-20}$ and a 95.5\% confidence interval of $[-16,0]\times10^{-20}$ for the stray-field dc Stark shift, in units of the clock frequency.

Briefly, we note that our group has recently developed techniques to efficiently manipulate the spatial distribution of lattice-trapped atoms in our OLCs~\cite{SieMcGHas24,HasKobBot24}. Such techniques can benefit clock operation. For instance, delocalized ensembles offer suppressed cold-collision (density) shifts without compromising atom number. In general, delocalized ensembles will be more susceptible to dc Stark shift analysis error attributed to field gradients. At the same time, these techniques could allow more rigorous constraints on the effects of field gradients by, e.g., loading the atoms in different locations within the shield.

Here we have demonstrated effective Faraday shielding and the ability of our shield to constrain residual stray-field dc Stark shifts. Tighter constraints than demonstrated here should be readily attainable, if deemed necessary for future clock operation.

\section{Thermo-mechanical design considerations}

The shield metal parts are computer numerical control (CNC) machined from electrolytic tough pitch (ETP) copper for high thermal conductivity. We chose ETP copper to build the shield structure to benefit from its lower ductility compared to the softer oxygen-free high thermal conductivity (OFHC) variant. Locking helical inserts and washers are used for most of the screw attachments to minimize loosening from thermal cycling. All copper surfaces were electro-polished to decrease their emissivities and reduce the thermal load from radiation, with the exception of the black-coated surfaces, which were treated by the coating vendor \cite{adibekyan2017high} to ensure specification-compliant coatings. Schott N-BK7 was chosen as the bulk material for the cryogenic optical windows attached to the shield instead of higher thermal conductivity optical substrates such as sapphire due to its superior absorbance of thermal radiation. For the shutter sphere mounting inside the stationary shield, we utilize all ceramic (ZrO$_2$) ball bearings with PTFE retaining rings, which require no lubrication and are ultra-high-vacuum (UHV) compatible. Several measures are implemented to facilitate the smooth and long-life operation of the ball bearings. First, the zirconium dioxide (ZrO$_2$) was chosen as the ceramic material due to its close coefficient of thermal expansion (CTE) to copper. The PTFE rings were modified by introducing an intentional radial cut to prevent clutching on the inner bearing race from contraction upon cooling to cryogenic temperatures. The bearings are fit with a custom-designed stainless steel ring to act as a radiation shield, minimizing ambient thermal radiation flux on the otherwise thermally isolated balls, thus ensuring uniform temperature of the bearings. The bearings are also specified at a loose radial play \cite{commercialSM}, and the housing in the copper shield is over-sized to accommodate any residual differential contraction between the two geometries. The loose tolerances are later remedied by utilizing wave-springs and retaining rings in the final assembly to pre-load the inner races of both bearings to the inward axial direction, providing an assembly that can accommodate differential thermal expansion, while still preserving the mechanical precision of the rotary motion of the shutter sphere. 

The shutter sphere is actuated by an out-of-vacuum stepper motor driven by a programmable stepper driver and triggered by the experimental sequence. The stepper motor performs a single rotational actuation of $45 \degree$ from one shield configuration to the other in $\sim$ 80~ms. The rotation is coupled into the UHV chamber through a magnetic rotary feedthrough, and a flexible beam coupler is used to compensate for any offsets or tilts of the shutter sphere axis and the chamber. The motor accelerates the shutter sphere for the first half of the actuation, then decelerates it for the second half, absorbing its rotational energy and minimizing overshoot and oscillations due to the large mass of the shutter sphere and the effective torsion spring from the magnetic and mechanical coupling of the feedthrough and the beam coupler, respectively. Regardless, we do observe oscillations with an amplitude of $< 0.5\degree$ peak to peak towards the end of the actuation that settle down completely in $<$ 100~ms. The BBR environment is minimally affected by the positional variation of the shutter sphere during the settling motion at the end of the actuation to the closed configuration, since all apertures are all adequately blocked well before the end of actuation. Still, we allow for 200~ms after the stepper driver trigger before performing any spectroscopy for reasons discussed later in this text, rendering the settling behavior even less consequential. The magnetic field produced by the magnets of the rotary feedthrough is estimated to be $<$ 1~mG (0.1 $\mu$T) at the atoms' location (assuming a magnetic dipole scaling from the manufacturer-specified residual field far from the magnets). We note that this field is constant during clock spectroscopy, and is stable shot-to-shot given the $< 0.1\degree$ reproducibility of the positional actuation of the shutter sphere over short and long timescales.

The three main shield body components are cooled through conductive coupling to an in-vacuum cold tip through a series of flexible copper braids \cite{commercialSM}. The braids are the thermal bottlenecks to the cold tip compared to the bulk conductivities of the copper and the thermal interface material (indium foil) between the braid and the shield part. Thus, the number of thermal braids to each part is roughly matched to the expected thermal load ratio on each component, and hence reduces the amount of required applied heat to equalize the shield components' temperatures. The cold tip is part of a commercial closed-cycle recirculating helium flow cryogenic system \cite{commercialSM}. The helium flow setup is designed to reduce vibrations from the pulsing Gifford-McMahon cryocooler head, which is mounted off the table that holds the clock optical and vacuum setup. The recirculating helium functions as a medium that transfers heat from the cold tip in the UHV chamber to the cryocooler head. High-purity helium is used and maintained by an inline gas purifier, as the helium temperature could get as low as 20 K in the coldest part of the flow. Contamination with heavier gas species deteriorates the performance of the cryogenic system when such gases freeze inside the flow lines, leading to restricted flow and reduced cooling power, or in extreme cases, a completely blocked helium flow. In both cases, warming up to room temperatures relieves the flow constriction or the clog. Initially, the cryocooler system provided reliable and stable baseline temperatures for more than 20 days in a test setup. Currently, the system provides usable cooling up to 4 days, after which a clog forms and a warm-up to room temperature is required. The vendor recommends regenerating the inline gas purifier to restore the long-term operation of the cryocooler, which is currently being explored as a solution.

Large magnetic fields are pulsed during the MOT loading phase, which lead to induced eddy currents in bulk conductive materials overlapping in space with the changing magnetic flux. We implement measures to break large circumferential conductive paths to mitigate the effect in the large copper parts of the shield. Electrical insulating film is introduced between the two shield halves, and a radial cut is intentionally introduced on the shutter sphere (see Fig.~\ref{fig:Temp long term}). Both measures are meant to eliminate the conductive paths going around the shield circumference in the horizontal plane, thus suppressing decaying magnetic fields in the vertical direction ($z$-axis) during the MOT loading phase. The measures are successful when the shield is at room temperature, as we observe no decaying magnetic fields during spectroscopy time. However, due to the higher electrical conductivity of copper at 77~K, we observe a non-negligible decaying field when spectroscopy starts soon after changing the applied magnetic field during atomic sample preparation phase. We measure a decay time constant of $\sim$~50 ms, and as a first-order measure wait at least 200 ms or 4 time constants for magnetic fields to stabilize before performing any sensitive spectroscopic measurements. We hypothesize that eddy currents in directions other than horizontal plane are produced, albeit smaller than those avoided by the measures described earlier. These currents potentially produce magnetic fields that are aligned along the magnetic field applied in the horizontal plane ($y$-axis) for clock spectroscopy. More cuts can be introduced in the shield design to eliminate such conductive loops, particularly in planes that are perpendicular to applied field during spectroscopy.

\section{Estimating the thermal load on the shield}
\label{sec: Thermal_load}
We estimate the thermal load on the shield by recording its temperature after the cryocooler is turned off and the shield is allowed to warm-up to room temperature. We assume that the shield is subjected to conductive and radiative thermal loads from a thermal reservoir at the room temperature, and neglect convective thermal loads since the shield is in a UHV chamber. The shield temperature in the warm-up period thus obeys the oridnary differential equation (ODE) model:
\begin{equation}
    mc_p(T)\frac{dT}{dt}=\sigma A_\mathrm{eff}(T_\mathrm{ext}^4-T^4)+k(T_\mathrm{ext}-T),
    \label{eq: heatload}
\end{equation}
where $m\sim2.5~\mathrm{kg}$ is the total mass of the shield and the copper braids, $c_p(T)$ is the copper specific heat capacity at constant pressure as a function of temperature in Ref \cite{white1984heat}, $\sigma=5.67\times 10^{-8}~\mathrm{Wm^{-2}K^{-4}}$ is the Stefan-Boltzmann constant, $A_\mathrm{eff}$ and $k$ are fitting constants for the radiative and conductive loads respectively, and $T_\mathrm{ext}=296$~K is the ambient temperature of the laboratory. This model assumes that all shield parts have the same temperature during the warm-up process, which is true for the stationary radiation shield halves ($<$ 200 mK difference during the warm-up process) but is not true for the shutter sphere or the braids (up to 10 K different from the radiation shield mean temperature). However, from thermal simulations and the observed thermal gradients on the shield parts, we estimate that more than 90\% of thermal load (conductive and radiative) is on the stationary radiation shield. Thus, we assign the temperature in the ODE model to be the mean temperature of the stationary radiation shield. The shutter sphere and the braids are treated as an extra thermal mass attached to the stationary shield, with the difference in temperature affecting the thermal evolution only through the temperature dependence the specific heat capacity. The difference in heat capacity due to 10~K difference in temperature of the shutter sphere or the braids is $\sim$ 15\% and is negligible for the estimating the total thermal load at the 5\% percent level. The solution of the ODE best fits the observed temperature evolution with $A_\mathrm{eff}=6.5(2)\times10^{-3} ~\mathrm{m^2}$ and $k=7.4(3)\times10^{-3}~\mathrm{W K^{-1}}$, with maximum deviation of the model temperature from the observed temperature $<1$ K over the warm-up period from 74~K to 295~K. The right-hand-side and left-hand-side of Eq.~(\ref{eq: heatload}) agrees at the few percent level between 74~K and 250~K, establishing confidence in the model. Plugging the numbers back into the ODE, we estimate a total heat load of 4.4 W at 77~K, with 2.8 W as radiative thermal load and 1.6 W as conductive contribution.  Finite element simulations show that the radiative thermal load gets similar contributions from the windows (emissivity $\approx$1) and the external surfaces (effective emissivity $\approx$0.24), whereas the conductive contribution is dominated by the stainless steel support screws and the PEEK washers that hold the shield to the room temperature vacuum chamber.

\section{Shield temperature stability}
The multimeter functions also as a temperature controller used in a closed-loop feedback to stabilize the temperature of the shield. As mentioned above, the number of copper braids between the cold tip and the different shield parts was chosen to be proportional to the calculated thermal load on each at cryogenic temperatures. Without thermal feedback, the realized baseline temperatures are $\approx$72~K and $\approx$71~K for the two shield halves and $\sim$ 67~K for the shutter sphere in the final installation in the clock apparatus. Thus, thermal feedback is necessary to equilibrate the main shield components seen by the atoms. We use thin film heaters attached to the shield side of the thermal braids. With a nominal heater power of $\sim$ 40~mW, $\sim$ 280~mW and $\sim$ 430~mW applied to the shield halves and the shutter sphere, we servo the temperature of the copper shield components to 77~K with a precision of less than a few mK. The thermal feedback loop senses the shield temperature through three of the 8 RTDs attached to the shield, one for each copper piece, positioned close to the copper thermal braids. Figure \ref{fig:Temp long term} shows the RTD locations in the shutter sphere, as well as stability plots for each of the in-loop and out-of-loop RTDs.
\begin{figure*}
\centering
\includegraphics[width=0.90\textwidth]{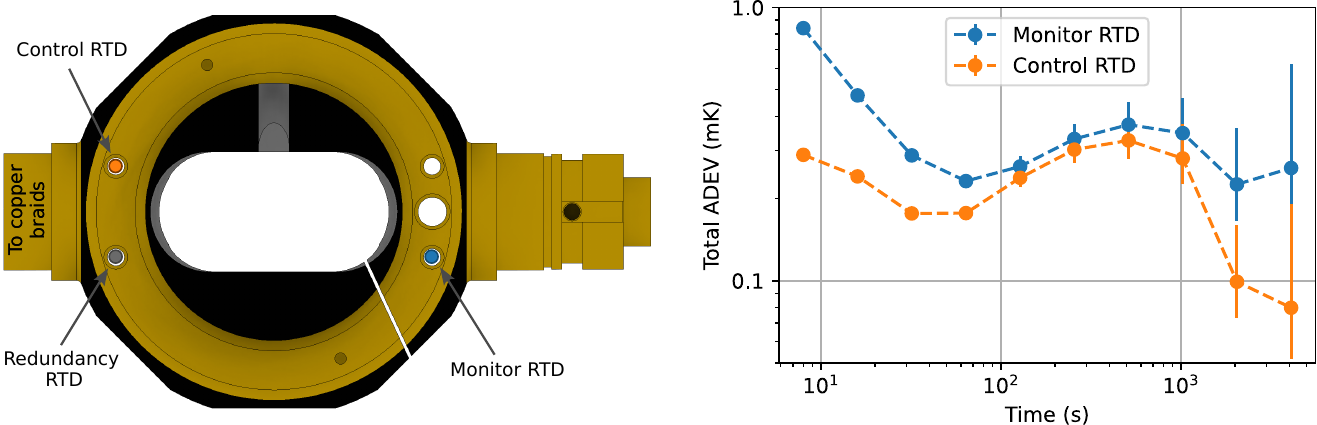}
\caption {(left) Location of the three RTDs embedded in the shutter sphere. Two  RTDs are placed close to the coldest part of the sphere, where the flexible copper braids to the cold tip are thermally attached. The two RTDs agree within the manufacturer uncertainty (5~mK). One of the two RTDs is used for the closed-loop feedback control (Control RTD). The third sensor (Monitor RTD) is mounted on the far-side from the copper braid. In addition to being used to measure the gradient across the shutter sphere, it also acts as an out of loop sensor for temperature stability. A black-coated circular copper cap covers the RTD region, and is not shown in the figure. (Right) Total Allan deviation of two sensor readings on the shutter sphere, shown for averaging time $\geq$ 8~s. Shorter averaging time is dominated by sensor noise.}
\label{fig:Temp long term}
\end{figure*}

\section{Temperature-dependent background gas collision bias in determining $\nu_\mathrm{dyn,6}$}
An important systematic bias that can be correlated with varying the shield temperature is the residual vacuum gas pressure. From residual gas analysis (RGA) of previous bake-outs of the shield and the Yb2 system at room temperature, we confirm that the dominant residual gas species in our vacuum system is hydrogen ($>$90\%).  As the temperature of the shield is varied, outgassing and pumping dynamics change, resulting in correlated variation of the residual background gas (BG) and possibly its composition. When the shield is operated at 77~K, it acts as a cryopump for many of the higher-mass gas species. RGA shows that the residual hydrogen becomes even more dominant than at room-temperature operation. It also shows that the partial pressure of all gas species including hydrogen is reduced by at least a factor of 2.5 at 77~K shield temperature. We thus anticipate that lower shield temperatures lead to less background gas collisional shift and more dominance of hydrogen species. On the contrary, operating the shield at elevated temperatures (318 K or 45 \degree C) can possibly lead to more outgassing than room temperature operation, causing an increased BG collisional shift. We note that the RGA doesn't indicate gas species other than those typically observed for a clean UHV system ($\mathrm{H_2O,\ N_2,\ CO_2,}$ etc.) up to 200 amu during bake-out at a temperature of 120 \degree C. 

We measure the atomic lattice trap lifetime as a proxy to estimate and correct for the BG shift, which follows the relation $(\Delta\nu/\nu)_\mathrm{BG}=-1.64(12)\times10^{-17}/\tau$ in \cite{McGZhaFas18}, where $\tau$ is the measured BG-limited lifetime in seconds. The Yb2 system employs an enhancement cavity for the lattice to access deeper traps with larger waist to reduce in-lattice atomic collision shifts. However, enhancement cavities can suffer from FM to AM noise conversion, leading to parametric heating and reduced trap lifetime, which complicates the estimation of BG shifts. By careful optimization of the cavity lock parameters, in addition to using a high-bandwidth ($\sim 500$ kHz) electro-optic modulator for the lattice intensity servo, we observed lattice lifetimes of 24.2(3)~s and 8.4(1)~s at 295~K and 318~K, respectively. We believe that these are fair estimates for BG-limited trap lifetimes at those temperatures (though they could still potentially be limited by residual parametric heating), and therefore we use them only to estimate upper bounds on the BG shift in the analysis below. We find that direct measurement of the trap lifetime at 77~K is more challenging. We have observed shorter lifetimes at these temperatures possibly due to extra vibrational noise that the cold-head exports to the out-of-vacuum optics, or to the shield through the pulsed helium flow in the cold tip. This requires additional study and analysis.

To account for the temperature dependence of BG shift on Yb2, possibly biasing the determination of $\nu_\mathrm{dyn,6}$ in the main paper, we consider two extreme cases: Case 0, with no variation of BG shift as a function of temperature, and case 1 with the maximum conceivable BG shift variation. We ascribe 25~s trap lifetime for all temperatures in case 0, and a conservative $4\times25\ \mathrm{s}=100$~s (motivated by the ideal gas scaling $P \propto T$), 25~s and 8~s for 77~K, 298~K and 318~K shield temperatures, respectively, in case 1. We apply the BG shift to the measurement runs of Yb2$-$Yb1 and Yb2$-$YbT and and extract $\nu_\mathrm{dyn,6}$ for each case. The difference between case 0 ($\nu_\mathrm{dyn,6}=-22.73(47)$~mHz) and case 1 ($\nu_\mathrm{dyn,6}=-22.20(47)$~mHz) is 0.53~mHz, which is only 12\% larger than the reported fit uncertainty. To properly account for this in the reported value, and given that the true slope could fall anywhere between the fitted slopes of case 0 and case 1, a uniform distribution is assumed for the true slope. The reported value in the main text is calculated as the mean of the two fitted slopes from the extreme cases. The uncertainty is determined by adding in quadrature the fitted slope uncertainty (0.475~mHz) and the standard deviation of the assumed distribution (0.162~mHz). As a result, the final reported value $\nu_\mathrm{dyn,6}=-22.47(50)$~mHz encompasses both extreme cases within 0.6$\sigma$.

We note that the measurements acquired at 120 K, 180 K, and 240 K in Fig.~2 in the main text are possibly impacted by an additional BG shift, since Yb2 shield temperature was ramped up to the temperature set points and care must be taken to make sure cryopumped gases released have had enough time to get pumped out. We argue that the effect is negligible for the three set points. For a typical passive 24 hr warm-up process to room temperature, a large burst of outgassing ($\sim 10\times$ the steady-state pressure level) happens between 77~K and 100~K, after which the released gas is pumped down with a time constant of approximately 1 hour. For the 120~K point, we stabilize the temperature for more than 12 hours before we start the comparison, which allows the gas to be pumped down to baseline levels before the measurement. We don't observe other gas bursts far above the baseline during the warm-up process up to room temperature, even close to 180~K and 240~K. For those set points, we don't wait as long before the measurement ($<$ 1 hour). To avoid any bias, we exclude those points from fitting the dynamic coefficient. RGA from previous test phases confirms that hydrogen gas is the dominant gas at all those temperatures.

\section{Literature values of the dynamic correction for $\mathrm{Sr}$ and $\mathrm{Yb}$}
Reported values of the dynamic corrections over the past 12 years for both $\mathrm{Sr}$ and $\mathrm{Yb}$ are shown in Fig \ref{fig:vdyn lit}. Over that period, the uncertainty on the dynamic correction has been a limiting source of uncertainty in optical lattice clocks that utilize either Sr or Yb. Conventionally, determining the dynamic correction involves theoretical atomic structure calculations informed by experimental measurements of various atomic properties such as state energies, lifetimes and branching ratios, magic and tune-out wavelengths. Notably, the lifetime measurements of the 5s4d \ls{3}{D}{1}{} state in Sr and the 6s5d \ls{3}{D}{1}{} state in Yb play a crucial role in determining $\nu_\mathrm{dyn,6}$. This is due to the relatively long \ls{3}{P}{0}{} $\leftrightarrow$ \ls{3}{D}{1}{} transition wavelength (2.6~$\mu$m in Sr and 1.4~$\mu$m in Yb) and large linewidth, leading to a large overlap of the non-static component of the differential clock polarizability with the room temperature BBR spectrum. Theoretical calculations of $\nu_\mathrm{dyn,6}$ are often limited by the precision of that single lifetime measurement. We refer to this conventional method of determining $\nu_\mathrm{dyn,6}$ as semi-empirical. Few exceptions to that method are \textit{ab initio} calculations performed for Yb in Ref.~\cite{safronova2012ytterbium} and direct experimental observation of the dynamic correction for Sr in Ref.~\cite{UshTakDas15}. We note the non-trivial evolution of the dynamic correction for Sr since 2015, where a re-evaluation disagrees with the previous value by $\sim 2\sigma$ on theoretical grounds or by $\sim 5\sigma$ due to experimental revision of the \ls{3}{D}{1}{} lifetime (here $\sigma$ refers to the 68\% confidence level of the evaluation). Such evolution led to re-evaluation of the clock accuracy at the $10^{-17}$ level over the course of 9 years. This underscores the importance of independent evaluations of the dynamic correction for optical lattice clock species. The present work helps establish confidence in $\nu_\mathrm{dyn,6}$ for Yb. Finally, this measurement is unaffected by the Taylor series expansion issues identified in Ref.~\cite{LisDorNos21}, as the expansion for Yb does not introduce the theory error observed in Sr.

\begin{figure*}
\centering
\includegraphics[width=1\textwidth]{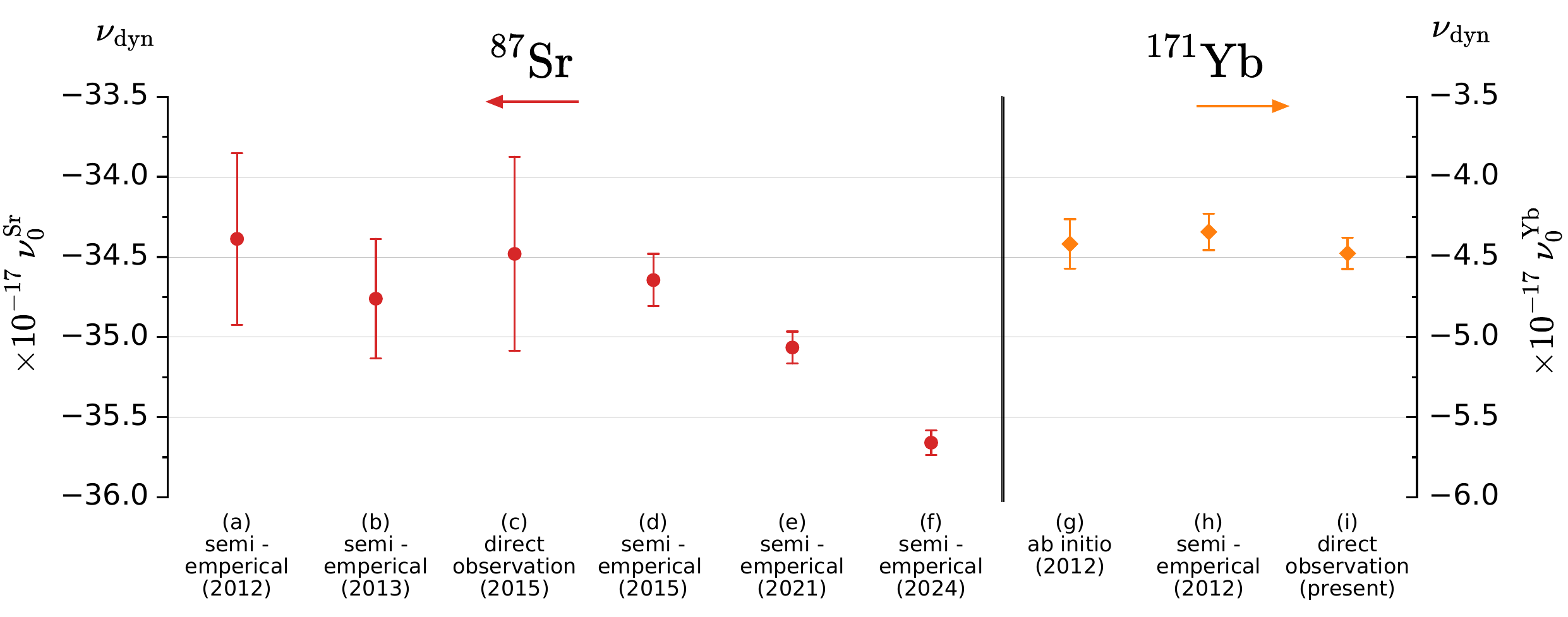}
\caption {Dynamic contribution to the BBR clock shift at 300~K for $^{87}$Sr and $^{171}$Yb in the literature over the past 12 years. The $^{87}$Sr data appear left of the vertical divider and are referenced to left axis (red); the $^{171}$Yb data appear right of the vertical divider and are referenced to the right axis (orange). The vertical scale is the same in units of the respective clock frequency. The values reported for Sr are (a) from \cite{MidFalLis12}, (b) from \cite{safronova2013blackbody} (c) from \cite{UshTakDas15} (d) from \cite{nicholson2015systematic} (e) from \cite{LisDorNos21} (f) from \cite{AepKimWar24}, while those for Yb (with $\nu_\mathrm{dyn}=\nu_\mathrm{dyn,6}+\nu_\mathrm{dyn,8}$) are (g) from \cite{safronova2012ytterbium} (h) from \cite{beloy2012determination} (i) with $\nu_\mathrm{dyn,6}$ from this work and $\nu_\mathrm{dyn,8}$ from \cite{beloy2012determination}.}
\label{fig:vdyn lit}
\end{figure*}

\end{widetext}

\begin{thebibliography}{10}
\bibitem{commercial}
The closed cycle cryocooler is a custom split-flow design based on the ColdEdge Stinger, specified at an estimated cooling power of 5~W at 50~K. Lakeshore provides the negative-temperature-coefficient RTDs (Cernox CX-1080-SD-HT) whose calibrations are NIST-traceable. The RTDs are paired with a Lakeshore model 336 cryogenic temperature controller for temperature readout and thermal feedback. Specification of commercial products is for informational purposes only. This does not imply recommendation or endorsement of any commercial product by NIST, nor does it imply that the products identified are necessarily the best available for the experimental purposes.
\bibitem{noisefloorpts}
Three data points reached the noise floor at {$\sim 5\times 10^{-18}$}. For these points, we assigned the floor level as the uncertainty.
\end{thebibliography}
\begin{thebibliography}{10}
\bibitem{commercialSM}
The closed-cycle cryocooler is a custom split-flow design based on the ColdEdge Stinger, specified at an estimated cooling power of 5~W at 50~K. Lakeshore provides the negative-temperature-coefficient RTDs (Cernox CX-1080-SD-HT) whose calibrations are NIST-traceable. The RTDs are paired with a Lakeshore model 336 cryogenic temperature controller for temperature readout and thermal feedback. The standard resistor used for transfer calibration (HARS-LX-2-100) is provided by IET labs with NIST-traceable calibration. Ceramic ball bearings are provided by SMB bearings, with custom C4 tolerance. Copper braids are provided by TAI and are custom designed to fit the interfaces on the shield and cold-tip. Specification of commercial products is for informational purposes only. This does not imply recommendation or endorsement of any commercial product by NIST, nor does it imply that the products identified are necessarily the best available for the experimental purposes.
\end{thebibliography}
\end{document}